\documentclass{aa}  
\usepackage{natbib}
\usepackage{graphicx}
\usepackage{sidecap}
\usepackage{float}
\usepackage[none]{hyphenat}
\usepackage[pdfpagelabels=false]{hyperref}
\hypersetup{colorlinks=true,linkcolor=blue,citecolor=blue,filecolor=blue,urlcolor=blue,}
\makeatletter
\renewcommand*\aa@pageof{, page \thepage{} of \pageref*{LastPage}}
\makeatother

\usepackage{adjustbox}  
\usepackage{amsmath}    
\usepackage{amssymb}    
\usepackage{multirow}   
\usepackage{array}       
\usepackage{mathtools}
\usepackage{comment} 
\usepackage{array, makecell} 
\usepackage{longtable}
\usepackage{caption}
\usepackage{subcaption} 
\usepackage{xcolor}
\usepackage{isotope}
\definecolor{green_comm}{RGB}{0,160,0}
\definecolor{orange}{RGB}{255,165,0}

\def\TiI{\hbox{\rm Ti\,$\scriptstyle\rm I$}}
\def\VI{\hbox{\rm V\,$\scriptstyle\rm I$}}
\def\CrI{\hbox{\rm Cr\,$\scriptstyle\rm I$}}
\def\FeI{\hbox{\rm Fe\,$\scriptstyle\rm I$}}
\def\NaI{\hbox{\rm Na\,$\scriptstyle\rm I$}}
\def\MgI{\hbox{\rm Mg\,$\scriptstyle\rm I$}}

\def\MnI{\hbox{\rm Mn\,$\scriptstyle\rm I$}}

\def\CaI{\hbox{\rm Ca\,$\scriptstyle\rm I$}}
\def\GeI{\hbox{\rm Ge\,$\scriptstyle\rm I$}}
\def\GaI{\hbox{\rm Ga\,$\scriptstyle\rm I$}}
\def\HI{\hbox{\rm H\,$\scriptstyle\rm I$}}

\def\PrI{\hbox{\rm Pr\,$\scriptstyle\rm I$}}
\def\KI{\hbox{\rm K\,$\scriptstyle\rm I$}}
\def\NI{\hbox{\rm N\,$\scriptstyle\rm I$}}
\def\NiI{\hbox{\rm Ni\,$\scriptstyle\rm I$}}
\def\SI{\hbox{\rm S\,$\scriptstyle\rm I$}}
\def\DyI{\hbox{\rm Dy\,$\scriptstyle\rm I$}}
\def\RbI{\hbox{\rm Rb\,$\scriptstyle\rm I$}}

\def\CaII{\hbox{\rm Ca\,$\scriptstyle\rm II$}}
\def\BaII{\hbox{\rm Ba\,$\scriptstyle\rm II$}}
\def\FeII{\hbox{\rm Fe\,$\scriptstyle\rm II$}}

\def\CrII{\hbox{\rm Cr\,$\scriptstyle\rm II$}}
\def\TmII{\hbox{\rm Tm\,$\scriptstyle\rm II$}}

\def\MgII{\hbox{\rm Mg\,$\scriptstyle\rm II$}}
\def\MnII{\hbox{\rm Mn\,$\scriptstyle\rm II$}}

\newcommand{\planet}{KELT-20b/MASCARA-2b}
\def\kms{$\mathrm{km\,s}^{-1}$}
      
\newcolumntype{P}[1]{>{\centering\arraybackslash}p{#1}}

\begin{document}

   \title{The GAPS programme at TNG}

   \subtitle{LXXVIII. Phase-resolved detection of multiple atomic species in the atmosphere of \planet}

   \author{F. Biassoni \inst{1,2} \thanks{Corresponding author: \email{federico.biassoni@inaf.it}} 
          \and
          F. Borsa \inst{1}
          \and
          L. Fossati \inst{3}
          \and
          M.C. D'Arpa \inst{4}
          \and
          A.S. Bonomo \inst{5}
          \and
          G. Guilluy \inst{5}
          \and
          T. Zingales \inst{6,7}
          \and
          D. Sicilia \inst{8}
          \and
          A. Sozzetti \inst{5}
          \and
          M. Rainer \inst{1}
          \and
          M. Basilicata \inst{1}
          \and
          I. Carleo \inst{5}
          \and
          R. Cosentino \inst{8,9}
          \and
          R. Claudi \inst{6,10}
          \and
          P. Giacobbe \inst{5}
          \and
          A.~Harutyunyan \inst{9}
          \and
          L. Malavolta \inst{6,7}
          \and
          L. Mancini \inst{5,11,12}
          }

   \institute{INAF -- Osservatorio Astronomico di Brera, via E. Bianchi 46, 23807 Merate, Italy
         \and
         Dipartimento di Scienza e Alta Tecnologia, Università degli Studi dell'Insubria, via Valleggio 11, 22100 Como, Italy
         \and
         Space Research Institute, Austrian Academy of Sciences, Schmiedlstrasse 6, 8042 Graz, Austria
         \and
         INAF -- Osservatorio Astronomico di Palermo, Piazza del Parlamento, 1, 90134 Palermo, Italy
         \and
         INAF -- Osservatorio Astrofisico di Torino, via Osservatorio 20, 10025 Pino Torinese, Italy
         \and
         INAF -- Osservatorio Astronomico di Padova, vicolo dell’Osservatorio 5, 35122 Padova, Italy
         \and 
         Dipartimento di Fisica e Astronomia, Università degli Studi di Padova, vicolo dell’Osservatorio 3, 35122 Padova, Italy
         \and
         INAF -- Osservatorio Astrofisico di Catania, Via S. Sofia 78, 95123 Catania, Italy
         \and
         Fundación Galileo Galilei-INAF, Rambla José Ana Fernandez Pérez 7, 38712 Breña Baja, Tenerife, Spain
         \and
         Dipartimento di Matematica e Fisica, Università Roma Tre, via della Vasca Navale, 84, Roma, 00146, Italy
         \and
         Dipartimento di Fisica, Università di Roma “Tor Vergata”, via della Ricerca Scientifica 1, 00133 Roma, Italy
         \and
         Max-Planck-Institut für Astronomy, Königstuhl 17, 69117 Heidelberg, Germany
         }

   \date{Received 5 May 2026 / Accepted dd August 2026}
 
  \abstract
   {Among the different classes of exoplanets, ultra-hot Jupiters represent prime laboratories for atmospheric characterisation. With an equilibrium temperature of $\sim$2300\,K, \planet \, (HD~185603~b) is a well-studied ultra-hot Jupiter, for which several atomic species have already been detected in both transmission and emission spectroscopy.
   Nevertheless, important open questions remain regarding the completeness of the chemical inventory and the interpretation of spurious aliasing signals that may mimic genuine atmospheric signatures. In addition, asymmetries in the detected signals across transits can encode valuable information on the spatial structure and dynamics of the planetary atmosphere.
   In this work, we analyse seven transits of \planet \, observed with the high-resolution spectrograph HARPS-N at the Telescopio Nazionale Galileo. We apply the cross-correlation function technique to a large set of atomic and ionic species, accounting for aliasing effects that can lead to false detections. We confirm previous detections of \HI \,, \NaI \,, \MgI \,, \CrI \,, \FeI \,, and \FeII \,, and tentatively detect \CrII. We further report new detections of \KI \,, \CaI \,, \VI \,, \MnI \,, and \BaII \,, and a tentative detection of \PrI \, in the planetary atmosphere.
   By performing a phase-resolved analysis and splitting the transit into pre- and post-mid-transit phases, we investigate longitudinal variations in the detected signals. For several species, we observe a double-peaked structure in the $K_{\rm p} - V_{\mathrm{sys}}$ maps, notably in \CaI \,, \VI \,, \CrI \,, and \FeI \,, consistent with absorption arising from two distinct atmospheric limbs of the planet.}

    
   \maketitle

\section{Introduction}

High-resolution transmission spectroscopy has emerged as a powerful tool for probing the composition and dynamics of exoplanetary atmospheres, particularly those of close-in orbit exoplanets \citep{Hoeijmakers_2018}. By resolving individual atomic, ionic, and molecular lines, this technique allows us not only detect chemical species but also measure their Doppler shifts, providing direct constraints on atmospheric winds, rotation, and circulation patterns \citep{Snellen_2010,Louden_2015,Ehrenreich_2020}.
Among all types of planets, ultra-hot Jupiters occupy a privileged position for atmospheric characterisation. Due to their proximity to their host stars, these planets are almost all tidally locked and subject to intense stellar irradiation, which results in heating and inflation of their atmospheres \citep[e.g.][]{Owen_2019,Biassoni_2023}. 
The short orbital separation accompanied by the typically high temperature of the host star, in addition to tidal locking, leads to the high equilibrium temperature in the atmosphere of ultra-hot Jupiters \citep[$T_{\rm eq} >$ 2200\,K;][]{Parmentier_2018}, causing strong winds from the day- to night-side and molecule dissociation. These exoplanets are widely regarded by the scientific community as ideal laboratories for atmospheric studies \citep{Rainer_2021}. Their high temperatures, inflated atmospheres, and large pressure scale heights ($H$) result in strong absorption signals during transits, making them particularly valuable for detailed atmospheric analysis \citep{Kempton_2018}.

KELT-20b \citep{Lund_2017}, also known as MASCARA-2b \citep{Talens_2018}, is an ultra-hot Jupiter with an equilibrium temperature of $T_{\rm eq} =$ 2260$\pm$50\,K orbiting the A2\,V type host star HD~185603 with a period of approximately 3.47 days \citep{Talens_2018}.
Photometric transit observations give a precise value of its radius of $R_p = 1.74 \, R_{\mathrm{Jup}}$ \citep{Lund_2017}. However, due to the high rotational velocity of the host star \citep[$v \, \sin{i} =$ 114\,\kms \,,][]{Talens_2018}, only an upper limit on the planetary mass could be derived, yielding a 3$\sigma$ constraint of $M_p < 3.51 \, M_{\rm Jup}$ \citep{Lund_2017}.

Due to its characteristics (Table \ref{tab:systemParameters} contains the KELT-20/MASCARA-2 system parameters), extended atmosphere \citep[up to $\sim$ 545 $H$, with $H$ $\sim$ 281 km;][]{Casasayas_2019}, and high equilibrium temperature, \planet \, has long been one of the most studied and analysed planets, both during transits and secondary eclipses.
Several studies have focused on the analysis of individual lines of elements present in its atmosphere, exploiting the high-resolution transmission spectroscopy at the position of lines such as H$\alpha$, H$\beta$, H$\gamma$, and H$\delta$ hydrogen Balmer lines, as well as the \NaI \, D doublet, the \CaII \, infrared triplet, the \CaII \, H\&K lines, and multiple \FeII \, lines \citep{Casasayas_2018,Casasayas_2019,Nugroho_2020,Fossati_Biassoni_2023,Stangret_2024}.

By applying the cross-correlation function (CCF) technique \citep{Snellen_2010,Brogi_2012} to multiple transits, additional chemical species were detected in the atmosphere of \planet \,, such as \NaI \,, \MgI \,, \CrI \,, \CrII \,, \FeI \,, and \FeII \,  \citep{hoeijmakers_2020kelt,Nugroho_2020,Lenhart_2025}. At the same time, some further species were detected on its day-side by applying the CCF to secondary eclipses such as \NI \,, \SI \,, \CrI \,, \FeII \,, \FeI \,,   and  \NiI \, \citep{Borsa_2022,Cont_2022_kelt20,Kasper_2023,Petz_2024}.

Despite the growing number of detections, several open questions remain regarding the completeness of the detected chemical inventory and the interpretation of spurious aliasing signals considered to be true detections. Furthermore, asymmetries in the detected signals recorded during transit events may encode valuable information on the spatial and dynamical structure of the atmosphere \citep[e.g.][]{Nugroho_2020}.
Phase-resolved high-resolution transmission spectroscopy provides a unique opportunity to disentangle contributions from different atmospheric limbs of the planet. By separating the transit into pre- and post-mid-transit phases, it is possible to investigate longitudinal asymmetries in both chemical composition and atmospheric dynamics \citep[e.g.][]{Espinoza_2024}.

In this work, we present a comprehensive high-resolution transmission spectroscopy analysis of \planet \, based on an extended dataset of HARPS-N observations. By applying the CCF to multiple atomic and ionic species and by splitting the observed transit data into pre- and post-mid-transit phases, we report new atmospheric detections and investigate the phase dependence of the detected signals. Before claiming any new detection, we performed a dedicated study on the aliasing problem affecting all atomic and ionic species. The paper is structured as follows. Sect. \ref{Observation} is dedicated to the description of our observations, while the telluric correction, the data analysis based on the CCF, the Doppler shadow correction, and the introduction of the aliasing removal method are described in Sect \ref{Data analysis}. The CCF and aliasing correction results are described in the Sect. \ref{result}, while Sect. \ref{conclusion} is reserved for the conclusions.

\begin{table}
\caption{adopted parameters of the KELT-20/MASCARA-2 system.
}

\renewcommand{\arraystretch}{1.35}
\resizebox{9cm}{5.2cm}{
\begin{tabular}{lccc}
\hline
\hline
Parameter & Symbol & Value & Unit \\
\hline
Stellar parameters & & &\\
\hline

Effective temperature $^{(a)}$ & $T_{\mathrm{eff}}$ & $8980^{+90}_{-130}$ & K \\

Stellar mass $^{(a)}$ & $M_{\star}$ & $1.89^{+0.06}_{-0.05}$ & $M_{\odot}$ \\

Stellar radius $^{(b)}$ & $R_{\star}$ & $1.57^{+0.06}_{-0.06}$ & $R_{\odot}$ \\

Spectral type $^{(a)}$ &  & A2\,V & \\

Projected rotational velocity $^{(a)}$ & $v_{\star}$ $\sin{i}$ & $114\pm3$ & \kms \\

\hline
Planetary parameters & & & \\
\hline

Planetary mass $^{(b)}$ & $M_{p}$ & $< 3.51$ & $M_{\rm Jup}$ \\
Planetary radius $^{(b)}$ & $R_{p}$ & $1.74^{+0.07}_{-0.07}$ & $R_{\rm Jup}$ \\
Equilibrium temperature $^{(a)}$ & $T_{\rm eq}$ & $2260 \pm 50$ & K \\

\hline
Orbital parameters & & & \\
\hline

Semi-major axis $^{(b)}$ & $a$ & $0.0542^{+0.0014}_{-0.0021}$ & au \\

Period $^{(a)}$ & $P$ & $3.474119^{+5e-6}_{-6e-6}$ & d \\

Epoch $^{(d)}$ & $T_0$ & $2457909.5875^{+0.0003}_{-0.0002}$ & BJD \\

Ingress/Egress duration $^{(b)}$ & $T_{23}$ & $0.47904^{+0.01920}_{-0.01848}$ & h \\ 

Transit duration $^{(c)}$ & $T_{14}$ & $3.5755^{+0.0218}_{0.0211}$ & h \\

Radial velocity amplitude $^{(e,f)}$ & $K_{\star}$ & 0.32251 & \kms \\

Systemic velocity $^{(c)}$ & $V_{\mathrm{sys}}$ & $-24.48\pm0.04$ & \kms \\

Impact parameter $^{(b)}$ & $b$ & $0.503^{+0.025}_{-0.028}$ & \\

Projected obliquity $^{(b)}$ & $\lambda$ & $3.4\pm2.1$ & deg \\

Inclination $^{(b)}$ & $i$ & $86.12^{+0.28}_{-0.27}$ & deg \\

Eccentricity $^{(a)}$ & $e$ & 0 (fixed) & \\
\hline

\end{tabular}
}
\tablefoot{Values are taken from (a) \citet{Talens_2018}; (b) \citet{Lund_2017}; (c) \citet{Rainer_2021}; (d) \citet{Hoeijmakers_2020}; (e) \citet{Casasayas_2019};
(f) Assuming the upper mass limit for the planet.}
\label{tab:systemParameters}
\end{table}

\begin{table*}
\centering
\caption{log of the transit observations of KELT-20b/MASCARA-2b  used in this work.}
\label{log_observation}
\renewcommand{\arraystretch}{1.2}
\resizebox{17.9cm}{1.7cm}{
\begin{tabular}{cccccccc}

\hline
\hline
Night \#&
Night date      &
Programme   &
PI &
$T_{\rm exp}$ [s] &
\# of spectra (Out/In) &
 <S/N>@ 5500\AA{}&
 Airmass (min/max)\\
\hline
1 &
16 August 2017        &
CAT17A-38  &
Rebolo&
200                 &
90 (33/57)          &
61               &
1.00/2.13 \\
2 &
12 July 2018         &
CAT18A-34  &
Casasayas-Barris&
200                 &
111 (56/55)         &
93               &
1.00/1.58    \\
3 &
19 July 2018        &
CAT18A-34  &
Casasayas-Barris&
300                 &
78 (39/39)          &
105              &
1.00/1.42    \\
4 &
26 August 2019         &
GAPS       &
Micela&
600                 &
30 (10/20)          &
164              &
1.00/2.09    \\
5 &
02 September 2019         &
GAPS       &
Micela&
 600                 &
29 (8/21)           &
176              &
1.00/1.55    \\
6 &
31 July 2022         &
GAPS       &
Micela&
600                 &
27 (10/17)           &
115               &
1.00/1.44    \\
7 &
05 July 2025 &
BRIDGES &
Borsa &
600 &
30 (10/20) &
130 &
1.00/1.21 \\
\hline
    \end{tabular}
    }
\end{table*}

\section{Observations} \label{Observation}

We analysed seven different transits in the visual band of the planet \planet \, with the high-resolution (R$\approx$115\,000) HARPS-N spectrograph ($3800-6900$\,\AA{}) \citep{Cosentino_2012}, which is located at Telescopio Nazionale Galileo (TNG), between 2017 and 2025. The first three transits (nights 16 August 2017, 12 July 2018, and 19 July 2018) were taken from the TNG archive, while the subsequent three nights (26 August 2019, 02 September 2019, and 31 July 2022) were acquired within the context of the atmospheric characterisation part of the GAPS (Global Architecture of Planetary Systems) programme \citep[e.g.][]{Covino_2013,Borsa_2019,Fossati_Biassoni_2023}. The final transit night (05 July 2025) was observed as part of GAPS large proposal BRIDGES (Building a Road to the In-Depth investiGation of Exoplanetary atmosphereS) programme A48TAC$\_$52, P.I. Borsa. A log of all observations is provided in Table~\ref{log_observation}.

Transits 2 and 3 (nights 12 July 2018 and 19 July 2018) suffered an issue with the atmospheric dispersion corrector \citep[further details are found in][]{Casasayas_2019} causing wavelength-dependent flux losses especially in the bluest part of the spectra. Furthermore, transits 2 and 6 (nights 12 July 2018 and 31 July 2022) presented a drop in the signal-to-noise ratio (S/N) just before the beginning and near the end of the transit (Fig. \ref{airmass_snr}, left panel). Given that we evaluated the S/N for each exposure near the peak of HARPS-N sensitivity, that is $\sim$ 5500\,\AA{} (echelle spectral order 46), and discarded all the exposures with a S/N lower than 50. This selection left to us 395 exposures, 229 of them during the transit (Table \ref{log_observation}).
To enhance the quality of the CCF analysis, the first five spectral orders and the last one were excluded from all exposures due to their low S/N, focusing on the wavelength range between $\sim 4001$ and $\sim 6832$\,\AA{} \citep{Biassoni_2024}.

\section{Data analysis}\label{Data analysis}

\subsection{Telluric correction}
For every exposure of each night, we corrected the HARPS-N 1D merged spectral orders (s1d) from the H$_2$O and O$_2$ telluric absorption lines using the {\fontfamily{pcr}\selectfont molecfit} software \citep{Smette_2015,Kausch_2015} following the approach described in \cite{Biassoni_2024}.
However, our data analysis based on the CCF was conducted on the unmerged echelle spectral orders (e2ds), rather than the merged s1d products, as the former allow for a more accurate continuum normalisation across the full spectral range of the spectrograph.
Therefore, for each exposure, we divided the e2ds spectral orders by the telluric transmission model retrieved from the s1d correction, after reinterpolating the telluric profile on the e2ds wavelengths, in order to remove the H$_2$O and O$_2$ absorption features from the unmerged spectral orders \citep[e.g.][]{Hoeijmakers_2020,Biassoni_2024}.
Working with the e2ds is preferable not only because it improves the continuum normalisation, but also because the spectra preserve the original wavelength calibration of each spectral order. In contrast, the s1d spectra are generated by merging the e2ds orders and interpolating them onto a common, evenly spaced wavelength grid (0.01\,\AA), which introduces a resampling that can slightly distort the line positions.
However, we corrected from the tellurics the s1d spectra, rather than the e2ds orders. This choice was motivated by practical and computational considerations. The s1d spectra provide a single, merged one-dimensional format that allows for a faster and more stable optimisation of the telluric model across the full spectral range.

\subsection{CCF}

Before performing the CCF analysis, we pre-processed the e2ds echelle spectral orders as in \cite{Biassoni_2024}, applying the following prescriptions.
For each observed night, the telluric-corrected e2ds spectral data were reorganised by order into separate matrices. Each matrix corresponds to a single spectral order and contains, along its rows, the individual exposures acquired during the night, while the columns represent the pixels within that order. This restructuring results in one matrix per spectral order, preserving the temporal information across exposures while isolating each order for subsequent analysis.
Since ground-based observations are subjected to flux variations over time due to the Earth's atmospheres, each exposure was normalised individually by dividing the flux values by the mean flux across the pixels of that exposure.
This process was repeated independently for each spectral order and for each night, maintaining a consistent normalisation and analysis procedure across all seven planetary-transit observations.

After these steps, we stacked all the previous matrices corresponding to different spectral orders along the pixel axis, building for each night a single matrix whose horizontal axis spans the entire wavelength range from 4001 to 6832\,\AA{}, and whose vertical axis still represents the number of exposures. This results in a set of one-dimensional spectra, $x(t)$, one per exposure, $t$, analogous in structure to the HARPS-N s1d spectra, but with a uniform continuum normalisation across exposures and preserving the original wavelengths of the echelle orders.
The reconstructed spectra, $x(t)$, were then used as input for the CCF analysis, which we performed separately for each night. We adopted the normalised CCF as in \citet{Biassoni_2024}, i.e. 
\begin{equation}\label{CCF_norm}
    {\rm CCF(v,t)} = \frac{\sum_{k}{x_k(t) \cdot T_k(v)}}{\sum_k T_k(v)} \,,
\end{equation}

\noindent
where $v$ is the velocity at which the template, $T$, is shifted and $k$ refers to the pixels in our reconstructed 1D spectrum $x(t)$. 
The templates were shifted in the range $-200$\,\kms $\leq v \leq +200$\,\kms, with steps of 1\,\kms \, and interpolated along the same spectrum axis.
The resulting CCF was Doppler shifted in the stellar reference frame, interpolating within the radial velocity (RV) range $-150$\,\kms $\leq v \leq +150$\,\kms \, with steps of 1\,\kms \,, and normalised by the temporal mean of the CCFs out-of-transit, in order to remove the stellar features. 
For a more accurate normalisation, we further divided the CCF map by the median values across the exposures. 
We exploited this prescription for each inspected template and for each night separately.

Given that \planet \, exhibits an inverted pressure–temperature profile \citep[e.g.][]{Borsa_2022,Fossati_Biassoni_2023}, its atmosphere is far from being isothermal. Although the true atmospheric structure is likely complex and non-isothermal, the CCF technique at high spectral resolution is predominantly sensitive to the relative positions and overall pattern of spectral lines, while being less sensitive to the continuum level and to the exact details of the vertical temperature profile. In this context, isothermal templates provide suitable matched filters for identifying the presence of specific species. It is therefore essential to perform the CCF analysis using templates generated at a range of temperatures in order to more adequately capture the thermal structure of the atmosphere.
Hence, we used standard isothermal atmospheric templates with temperatures of 2000\,K, 3000\,K, 4000\,K, and 5000\,K provided by \cite{Kitzmann_2023} after convolving them with the HARPS-N instrumental profile (R $\approx$ 115\,000). These templates are computed under local thermodynamic equilibrium assuming a solar chemical abundance \citep{Kitzmann_2023}. Further information about opacities was provided by \cite{Kitzmann_2023}.
We investigated 81 neutral and ionised species at 2000\,K, 102 at 3000\,K, 111 at 4000\,K, and 114 at 5000\,K for each night separately.
We adopted all these templates since using the highest or lower temperature template for all ionic or atomic species is suboptimal. While a higher-temperature template may contain more lines for ions, and a lower-temperature one includes more transitions for neutral species, many of these additional lines can be physically weak or entirely absent in the actual data. Including these faint or missing lines in the CCF introduces numerical noise into the calculation, diluting the true planetary signal and degrading the S/N.
This temperature scanning procedure allows us to maximise the match between the model templates and the atmospheric signal of each chemical species. Such an approach enables us to detect species that might remain undetected when using only one or two template temperatures, as the strength and number of absorption lines vary with temperature, affecting the CCF signal. 
While the planet’s equilibrium temperature is $\sim 2260$\,K, higher-temperature templates can still probe hotter regions of the atmosphere, providing additional information on the distribution of elements across different atmospheric layers.

\subsection{Doppler shadow correction}\label{Doppler shadow correction}

A planetary transit produces a distortion in the stellar absorption lines, causing a time-dependent feature in the CCF maps, commonly known as Doppler shadow \citep[e.g.][]{hoeijmakers_2020kelt}. The shape and intensity of the Doppler shadow are influenced by the projected stellar rotational velocity, the projected stellar spin-orbit inclination, the impact parameter, the $R_p/R_{\star}$ ratio, and depend on the inspected element.

We modelled the RVs of the Doppler shadow ($v_{\mathrm{DS}}$) as in \cite{Rainer_2021}. These RVs can be estimated as
\begin{equation}\label{eq:vds}
     v_{\mathrm{DS}} = v_\star \sin{i} \, (x_p \cos{\lambda} - y_p \sin{\lambda}) \,,
\end{equation}

\noindent
with $x_p = a_{R_{\star}} \sin{(2 \pi \phi)}$ and $y_p = - a_{R_{\star}} \cos{(2 \pi \phi)} \, \cos{i}$, where $a_{R_{\star}}$ is the semi-major axis in units of stellar radius, $\lambda$ is the projected obliquity, $i$ is the orbital inclination, $v_\star \sin{i}$ the projected rotation speed, and $\phi$ the orbital phase. In Eq. (\ref{eq:vds}) the $\lambda$ and $i$ values must be given in radians.

Many of our CCFs show the characteristic Doppler shadow, particularly for those species with strong absorption lines in the stellar photosphere. For these elements, we corrected and removed the Doppler shadow for each individual night separately, fitting two Gaussians, one in absorption and one in emission (a double Gaussian profile) to each exposure (i.e. each phase in the CCF), centred on the Doppler shadow RVs according to Eq. \ref{eq:vds} \citep[e.g.][]{Bourrier_2018,Hoeijmakers_2020,Rainer_2021}. Figure \ref{DS} shows the CCF obtained with the \FeI \, isothermal template at 2000\,K for the first night before (left) and after (right) the Doppler shadow correction.
The Doppler shadow correction is crucial to avoid misinterpreting the planetary signals in the CCF maps.
The Doppler shadow correction was performed only for elements where it could be robustly fitted. For cases where the fit was not statistically significant, the Doppler shadow correction was not applied, preventing the addition of artificial noise to the data.

\subsection{Merged CCF} \label{Merged CCF}

Once all the CCFs for each species and night were obtained, and the Doppler shadow was corrected for each night, we built a single consolidated CCF for each species and temperature profile. To do this, we gathered all individual exposures (phases) from all seven nights and sorted them globally in ascending order of their orbital phases.
To improve the normalisation of the CCF map, we applied a Fourier filter to cut off all velocities larger than 200\,\kms. Additionally, to eliminate any stellar, telluric residuals, and any other sources of systematic noise constant in time, we fitted and divided a first-degree polynomial for each column (RV) of the two-dimensional CCF as in \cite{Prinoth_2022}. Finally, we set the CCF continuum equal to zero by subtracting 1.

The planetary trace has a time dependent RV according to the relation $v_p = K_p \sin{(2 \pi \phi)}$ (dashed red line in Fig. \ref{DS}, drew only for the out-of-transit spectra to avoid overlapping with the in-transit planetary signal), where, assuming a circular orbit, $M_p \ll M_{\star}$ with the planetary semi-amplitude $K_p = K_{\star} M_{\star}/M_p$ with $K_{\star}$ is the stellar RV amplitude, while $M_{\star}$ and $M_p$ are the stellar and planetary masses, respectively (Table \ref{tab:systemParameters} contains the used values). 
To maximise the signal from the planet, the reference frame of each exposure is shifted by applying a Doppler correction equal to the planet's velocity $v_p$. Repeating this process for different $K_{\rm p}$ values and averaging all the in-transit CCF values allows us to determine the $K_{\rm p}$ where the signal peaks. This construction leads to the so-called $K_{\rm p} - V_{\mathrm{sys}}$ map \citep{Brogi_2012}. 
For each of our merged CCF we constructed the $K_{\rm p} - V_{\mathrm{sys}}$ map, evaluating the $K_{\rm p}$ value between 0\,\kms \, and 350\,\kms \, with 1\,\kms \, of step. To give the $K_{\rm p}  - V_{\mathrm{sys}}$ maps in term of the S/N, we calculated their standard deviation $\Sigma$ after excluding a window with RV in the range $[-35$\,\kms, $+35$\,\kms $]$ and divided all the $K_{\rm p}  - V_{\mathrm{sys}}$ maps by their standard deviation.
During the analysis we took into account the systemic velocity and the barycentric Earth RV, and hence the planetary signal is expected to be centred at RV$=0$\,\kms.

Mathematically, the CCF defined in Eq. \ref{CCF_norm} is constructed such that the planetary contribution appears as an absorption feature. However, to enhance visual clarity and ensure that the detection significance is represented intuitively, we inverted the CCFs by multiplying them by $-1$. This transformation ensures that the planetary signal appears as a positive feature in both the CCF and $K_{\rm p}  - V_{\mathrm{sys}}$ maps, resulting in a positive S/N value. This approach facilitates the interpretation of the results while maintaining physical consistency with the cross-correlation process.

\begin{figure*}
    \centering
    \includegraphics[width=0.85\linewidth]{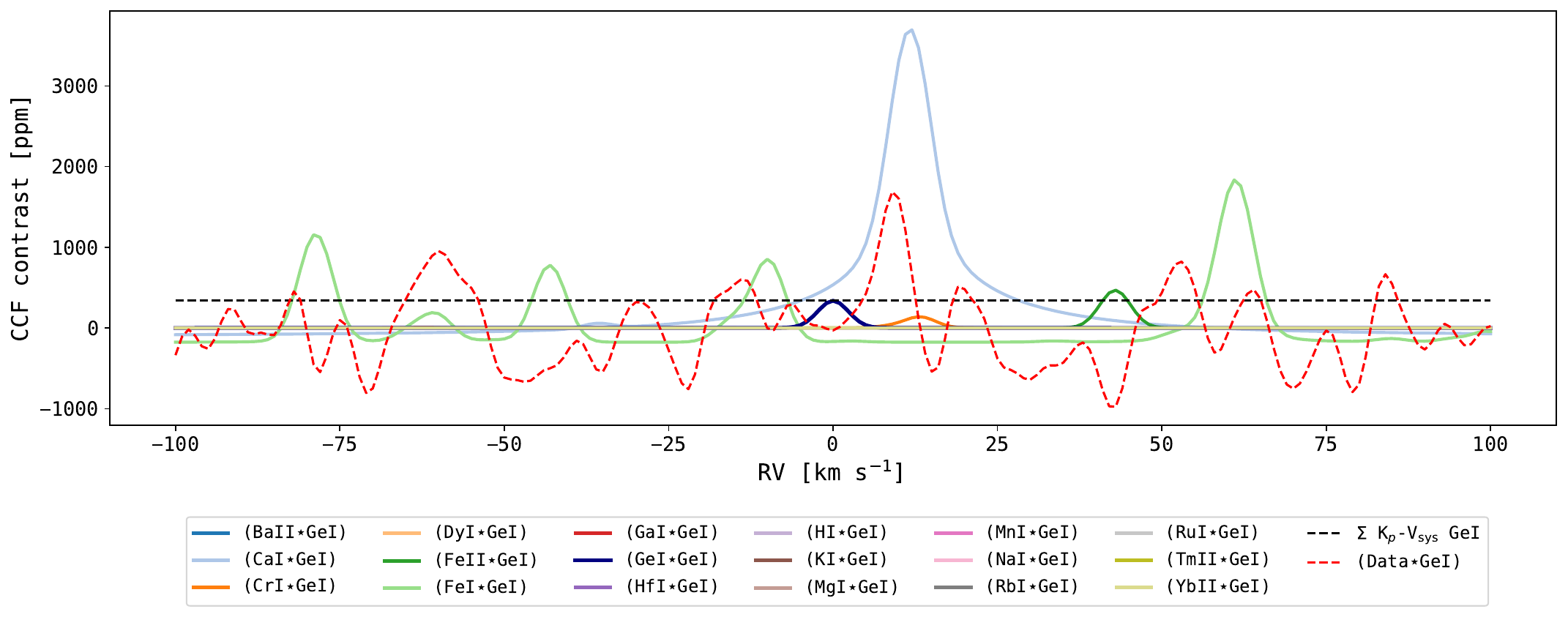}
    \caption{$K_{\rm p}  - V_{\mathrm{sys}}$ 1D curves from the $( e_i \,\star$ \GeI \,$)$ CCF at 5000\,K. The dashed red line is the 1D curve from the $K_{\rm p}  - V_{\mathrm{sys}}$ map of \GeI \, with observations, while the dashed horizontal black line is the standard deviation $\Sigma$ value of its $K_{\rm p}  - V_{\mathrm{sys}}$ map. The bold navy line is the autocorrelation (\GeI \,$\star$ \GeI \,)$_{\rm 1D}$.}
    \label{fig:aliasing_1D}
\end{figure*}

\subsection{Detection criterion}\label{Detection criterion}

Given the large number of analysed elements, we established a general criterion for identifying detections and non-detections. We decided to consider a detection in our $K_{\rm p}  - V_{\mathrm{sys}}$ maps any signal with a S/N $\geq 4$ \citep[as in][]{Biassoni_2024} in the $K_{\rm p}$ range between 100\,\kms \, and 250\,\kms \, and within a range of $\pm$10\,\kms \, centred at RV $= 0$\,\kms . We considered a tentative detection any signal with $3.5 < \mathrm{S/N} < 4.0$ with the same RV and $K_{\rm p}$ boundary conditions. 
By cutting off at velocities $> +10$\,\kms, we exclude any signals that could be spurious, spikes caused by the CCF, and that do not represent a genuine detection.
On the other hand, the limit at $> -10$\,\kms \, is applied because wind speeds on the order of $1 - 10$\,\kms \, are considered the range of velocities expected at the terminator region \citep{Miller_2012}. From Table \ref{tab:systemParameters}, we obtained a theoretical $K_{\rm p}$ value of $\sim 181.9$\,\kms. Hence, these $K_{\rm p}$ and RV limits help to avoid including noise or overly extreme RV blue-shifted signals that may not be reliable detections.

\subsection{Aliasing correction method}\label{Aliasing correction method}

It may happen that several observed signals that passed our detection criteria are not true detection. In other words, they are not signals from the element being investigated but rather spurious signals due to noise or the result of aliasing.
Aliasing occurs when lines from one template correlate with lines from other species. This can appear in the CCF and $K_{\rm p}  - V_{\mathrm{sys}}$ maps as a signal offset from the location where the actual planetary signal for the investigated species is expected to be. This issue is especially prevalent with templates that have only a few weak lines \citep{Borsato_2023}.
A major problem arises when these aliasing signals occur close in RV to the expected signal, as they can mimic the signal of the sought element, leading to misidentification and false detection.
Therefore, before confirming any results, it is essential to consider the possibility that the observed signal is the real detection of the element under investigation, rather than a false aliasing signal.

To determine whether the detected or tentative signals are real or the result of aliasing, we applied the following procedure, performed independently for each investigated temperature through the following steps.
First, let $\{e_i\}$ be the set of all detected and tentatively detected species at a given temperature, with $i = 1,...,N_{\theta}$ and $N_{\theta}$ the number of detections obtained at the temperature $\theta$. For each element $A = e_j$ in this set, we extract the $K_{\rm p}^{\mathrm{max}} (A)$ value from its observed $K_{\rm p}  - V_{\mathrm{sys}}$ map. This value corresponds to the $K_{\rm p}$ velocity where the maximum value of the $K_{\rm p}  - V_{\mathrm{sys}}$ map is found in a region between $-10$\,\kms $\leq$ RV $\leq +10$\,\kms \, and 100\,\kms $\leq$ $K_{\rm p}$ $\leq$ 250\,\kms.
Second, for every species $e_i$ (including $A$ itself), we build a 2D matrix, whose columns correspond to wavelengths and rows to all in-transit phases.
Each matrix contains the template spectrum of the element $e_i$ Doppler shifted in wavelength to match the planetary velocity, $v_p$, assuming $K_{\rm p} =$ $K_{\rm p}^{\mathrm{max}} (A)$.
Third, for each in-transit phase (between the $T_1$ and $T_4$ contact points), we computed a cross-correlation between the template of element $A$ and the 2D matrix of element $e_i$ using Eq. \ref{CCF_norm}, where $A$ acts as the normalised template $T$ and $e_i$ plays the role of observed data. Out-of-transit phases are added by setting the CCF values to zero.
This yields a synthetic CCF, noted as $(e_i \,\star \,A)$, where the entire notation $( f \, \star \, g)$ formally represents the CCF between two functions, $f$ and $g$. This matrix has the same dimension as the combined multi-night observed CCFs, where $e_i$ mimics the planetary signal at $K_{\rm p}^{\mathrm{max}} (A)$ and $A$ is the inspected template.

From this synthetic CCF, we computed the corresponding $K_{\rm p}  - V_{\mathrm{sys}}$ map and noted it as $(e_i \,\star \,A)_{K_p - V_{\mathrm{sys}}}$. These $(e_i \,\star \,A)_{K_p - V_{\mathrm{sys}}}$ maps are not scaled by their standard deviation $\Sigma$, instead they are represented by their numerical output similar to \cite{Borsato_2023}.
The templates provided by \cite{Kitzmann_2023} were derived using the Sun’s radius $R_{\odot}$ and a planetary radius of $1.5 R_{\mathrm{Jup}}$ as a reference. Due to the commutative and distributive properties of CCF and $K_{\rm p}  - V_{\mathrm{sys}}$ maps, respectively, we rescaled the obtained $(e_i \,\star \,A)_{K_p - V_{\mathrm{sys}}}$ maps by multiplying them by the factor
\begin{equation}
    \left( \frac{R_p / R_{\star}}{1.5 R_{\mathrm{Jup}} / 1 R_{\odot}} \right)^2 \,, 
\end{equation}
using $R_{\star}$ and $R_p$ from Table \ref{tab:systemParameters}.

Figure \ref{fig:aliasing} illustrates this process for templates at 5000\,K, where the inspected element $A$ is \GeI \, and the \CaI \, mimics the planetary signal, i.e. $( \CaI \,\star\, \GeI \,)$. It is visible the \CaI \, aliasing on the \GeI \, element during the CCF analysis in proximity to the expected RV planetary signal, highlighted by the light blue arrows in the $K_{\rm p}  - V_{\mathrm{sys}}$ map. Hence, we must be very careful in similar situations as the possible signal of \GeI \, may be a false signal due to \CaI \, aliasing. Since the \CaI \, signal was Doppler shifted assuming $K_{\rm p}^{\mathrm{max}} (A = \GeI) =$ 236\,\kms \, obtained from the \GeI \, $K_{\rm p}  - V_{\mathrm{sys}}$ map with data at 5000\,K, the maximum value of the $( \CaI \,\star\, \GeI \,)_{K_p - V_{\mathrm{sys}}}$ aliasing map indeed occurs at $K_{\rm p} =$ $K_{\rm p}^{\mathrm{max}} (A) =$ 236\,\kms. 

Once all the $K_{\rm p}  - V_{\mathrm{sys}}$ maps $(e_i \,\star \,A)_{K_p - V_{\mathrm{sys}}}$ for the element $A$ were generated at a given temperature $\theta$, we extracted from each of them the corresponding $K_{\rm p}  - V_{\mathrm{sys}}$ 1D curve at $K_{\rm p} =$ $K_{\rm p}^{\mathrm{max}} (A)$ noted as $(e_i \,\star \,A)_{\rm 1D}$, and subtracted their median value to ensure a common continuum level. These synthetic 1D curves, including the autocorrelation $(A \,\star \,A)_{\rm 1D}$, were compared with the observed curve $(Data \,\star \,A)_{\rm 1D}$ (Fig. \ref{fig:aliasing_1D}).
We computed the maximum value of each synthetic $(e_i \,\star \,A)_{\rm 1D}$ curve within $\pm$ 10\,\kms around RV $= 0$. We then applied a threshold selection: if the maximum of a synthetic curve exceeded the standard deviation $\Sigma$ of the observed data, it was flagged as a potential source of aliasing; otherwise, it was discarded as statistically insignificant.

The 1D curves that passed the $\Sigma$ threshold were then used as regressors in a multiple regression analysis aimed at modelling and removing the aliasing contribution from the observed $(Data \,\star \,A)$ signal.
Specifically, we adopted a Ridge regression (Tikhonov regularisation) implemented through the {\fontfamily{pcr}\selectfont Python RidgeCV} class of the \textit{scikit-learn} package \citep{Pedregosa_2011} minimising a penalised sum of squares:
\begin{equation}\label{ridge}
    \min_{w} ||Xw - y||_2^2 + \alpha||w||_2^2 \,\,,
\end{equation}

\noindent
where $w$ is the vector of weights, $X$ is the matrix whose columns are the synthetic $(e_k \,\star \,A)_{\rm 1D}$ curves, and $y$ is the observed signal $(Data \,\star \,A)_{\rm 1D}$. Since all the $K_{\rm p}  - V_{\mathrm{sys}}$ 1D curves are at the same continuum level, we did not fit an intercept in the regression. The regularisation strength $\alpha$ is automatically optimised via internal cross-validation over a logarithmic grid of values.
The regression was conducted only in a range of $\pm 25$\,\kms \, centred at RV $=$ 0 to avoid an overfit of possible spurious signals away from the detection, producing wrong weights. We selected this RV range to account for the broad profile of the curves in order to perfectly match the shape of the detection (Fig. \ref{fig:aliasing_1D}). The only exception is made for \BaII, where we relax this window to $\pm 50$\,\kms \, to accommodate its extremely wide profile. We employed Tikhonov regularisation rather than a simple multiple linear regression to obtain a stable solution in the presence of correlated predictors. This approach reduces the impact of noise-driven coefficients and minimises the risk of overfitting.

Once the vector of weights, $w$, was determined via the regression, they are used to perform the data correction. We multiplied these weights by their corresponding synthetic CCF, i.e. $w_k \cdot CCF_{k} = w_k \cdot (e_k \,\star \,A)$, where $k$ represents all the elements that overcome the $\Sigma$ threshold. Hence, we calculated the corrected CCF$_{\rm corr}^A$ of the inspected element $A$ subtracting the sum over all elements $k$ excluding the autocorrelation from the observed CCF of the element $A$,
\begin{equation}
\begin{aligned}
    CCF_{\rm corr}^A 
    &= \, CCF^A \, - \, \sum_{k \neq A} w_k \cdot CCF_k \, = \\
    &= \, (Data \,\star \,A) - \sum_{k \neq A} w_k \cdot (e_k \,\star \,A) \,\,.
\end{aligned}
\end{equation}

\noindent
We repeated this aliasing correction procedure for each of the detected and tentative detected elements of the whole set $\{e_i\}$.

As a final methodological remark, we note that in some cases, strong aliasing features can generate additional Doppler shadow signatures at the RVs corresponding to the alias signal. For these elements, we corrected the Doppler shadows prior to the aliasing correction. This was performed using the same procedure described in section \ref{Doppler shadow correction}, but constraining the double Gaussian fit to be centred at the RV of the Doppler shadow associated with the alias.

\section{Results and discussion}\label{result}

\begin{figure*}[ht!]
\centering
    \includegraphics[width=0.98\linewidth]{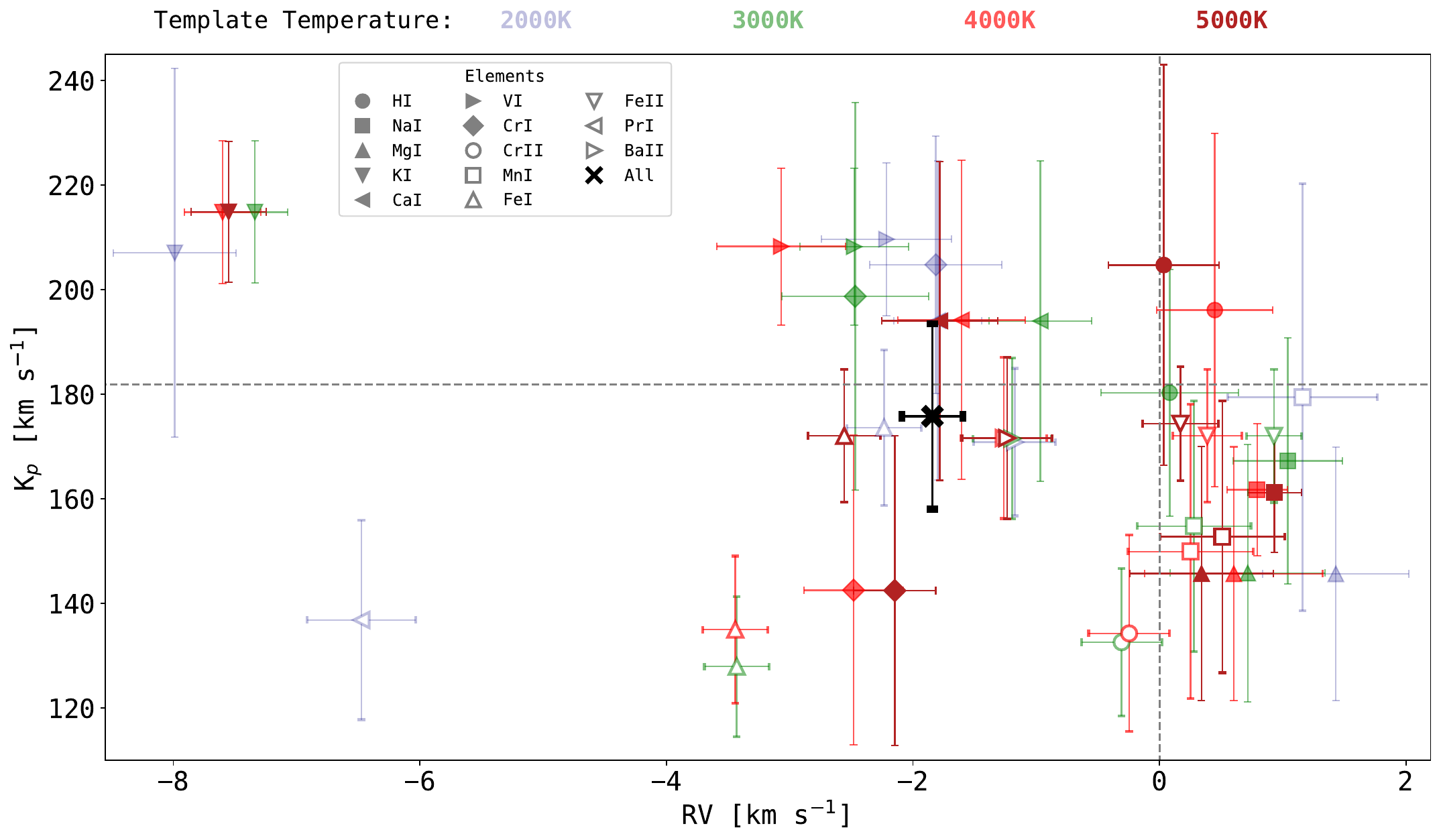}
    \caption{RV and $K_{\rm p}$ positions obtained from the Gaussian fit and the bootstrap approach, respectively, of our detected and tentative detected elements after the aliasing correction. The vertical and horizontal dashed grey lines mark our reference frame, i.e. RV = 0 and the theoretical semi-amplitude planetary orbital velocity $K_{\rm p}$. Each element is represented by different marker, while colours identify the temperature. We report also the values obtained with all the templates marked with a bold cross.}
    \label{fig:CCF_result}
\end{figure*}

The application of the CCF to the full dataset yields a number of significant signals for several atomic and ionic species. We employed templates at multiple temperatures to maximise the number of detectable species, since templates of the same element at different temperatures exhibit different line strengths and weights. Therefore, performing the CCF analysis across a range of temperatures ensures that potential signals are not missed and that detections are achieved with maximal sensitivity.

Before applying the aliasing correction, according to our detection criterion (Sect. \ref{Detection criterion}), we obtained from the associated $K_{\rm p}  - V_{\mathrm{sys}}$ for each species expressed in terms of S/N, a signal statistically significant (S/N $> 3.5$) for \MgI \,, \KI \,, \CaI \,, \CrI \,, \MnI \,, \FeI \,, \GaI \,, \GeI \,, \BaII \,, \DyI \,, and \TmII \, at all the inspected temperatures. A \VI \, signal was observed instead at 2000\,K, 3000\,K, and 4000\,K, while \HI \,, \NaI \,, \FeII \,, and \RbI \, were observed at 3000\,K, 4000\,K, and 5000\,K. At 3000\,K and 4000\,K we tentatively detected \CrII. Furthermore, \PrI \, was tentatively detected at only 2000\,K.
All these candidate detections were subsequently analysed using the aliasing correction procedure described in Sect. \ref{Aliasing correction method}. After accounting for aliasing spurious correlations, only a subset of these elements can be considered robust atmospheric detections.
Hence, we confirmed literature detection of \HI \,, \NaI \,, \MgI \,, \CrI \,, \FeI \,, and \FeII \,, and tentative detection of \CrII \,, \citep{hoeijmakers_2020kelt,Nugroho_2020,Lenhart_2025} and detected for the first time in the \planet \, atmosphere \KI \,, \CaI \,, \VI \,, \MnI \,, \BaII \,, and tentatively detected \PrI. 

For each of these detected and tentatively detected species that survived the aliasing correction, we derived the peak position values and uncertainties for RV and $K_{\rm p}$ by extracting one-dimensional curves from the $K_{\rm p}  - V_{\mathrm{sys}}$ maps. To derive the RV position, we extracted a horizontal 1D curve at the fixed $K_{\rm p}$ value where the planetary signal peaks ($K_{\rm p}^{\rm max}$) and fitted it with a Gaussian profile centred at RV $=$ 0 plus a constant offset \citep[similar to][]{Hoeijmakers_2019,Borsato_2023}. 
This approach is justified by the fact that the $K_{\rm p}  - V_{\mathrm{sys}}$ signal, when evaluated along the RV axis, is well approximated by a Gaussian profile. From the fit we derived the centre, amplitude (expressed in terms of S/N), and the width of the Gaussian. 
In contrast, the vertical 1D profile extracted along the $K_{\rm p}$ axis does not generally exhibit a Gaussian shape. Adopting a parametric fit here would introduce systematic biases. Therefore we employed a bootstrap approach as follows.
We extracted the 1D vertical profile along the $K_{\rm p}$ axis at the fixed RV bin corresponding to the maximum signal of the map found in the window between $-10$\,\kms $\leq$ RV $\leq +10$\,\kms \, as stated in Sect. \ref{Aliasing correction method}. Since the $K_{\rm p} - V_{\mathrm{sys}}$ maps are normalised by their standard deviation $\Sigma$ away from the signal (Sect. \ref{Merged CCF}), each point along this 1D profile is expressed in terms of S/N with an uncertainty equal to unity.
We generated $10^5$ realisations of this vertical profile by randomly sampling the S/N value of each $K_{\rm p}$ bin from a Gaussian distribution centred on its measured S/N where the standard deviation of the Gaussian function is set to $\sigma = 1$, representing an uncertainty of one dimensionless unit of S/N.
For each individual realisation, we identified the $K_{\rm p}$ position where the resampled signal reaches its maximum. The final $K_{\rm p}$ value was then defined as the mean of the resulting distribution of maxima, while its uncertainty was determined from the corresponding standard deviation. This bootstrap approach and the Gaussian fit gave consistent results when applied to the horizontal $K_{\rm p} - V_{\mathrm{sys}}$ 1D curves. However, the Gaussian fit was preferred for the RV axis as it simultaneously provides the signal amplitude.

Figure \ref{fig:CCF_result} shows all the species that passed the detection criterion threshold after the aliasing correction. The positions of their RV was found by the Gaussian fit while their $K_{\rm p}$ values were obtained from the bootstrap.
The vast majority of them lie in a narrow RV region between $-4$\,\kms \, and $+1$\,\kms. A weighted mean of the RVs position gives as result a global blue-shift of RV $\sim$ $-1.71 \pm 0.05$\,\kms \,, suggesting a global day-to-night side wind in the whole terminator region. Meanwhile the average $K_{\rm p}$ value is slightly below the theoretical one $K_{\rm p} \sim 173.1 \pm 2.7$\,\kms \, suggesting an asymmetric and expanded day side atmospheric geometry if compared with the planetary night side \citep[][]{Prinoth_2022}.
Figure \ref{fig:CCF_result} shows that many species are distributed over a wide range of $K_{\rm p}$ values, even for the same element at different temperatures. This suggests that the detected species probe different atmospheric layers, and that by varying the temperature of the templates we become sensitive to different regions of the planetary atmosphere. 
Overall, the retrieved points span a broad range in $K_{\rm p}$, extending from $\sim 128$\,\kms \, to $\sim 214$\,\kms.
The observed spread in $K_{\rm p}$ may be driven by atmospheric dynamical effects. However, when considering the associated uncertainties, all measurements remain consistent within 3$\sigma$ with the theoretical $K_{\rm p}$ value. A similar $K_{\rm p}$ spread have already been reported for other ultra-hot Jupiters, such as KELT-9b \citep{Borsato_2023} or WASP-76b \citep{Kesseli_2022}, supporting the interpretation of a stratified and dynamically complex atmosphere.
It can also be noted that some species (e.g. \CrI \, and \FeI) exhibit different $K_{\rm p}$ values depending on the adopted temperature. This behaviour is driven by spatial and thermal asymmetries between the planet’s morning and evening limbs. For globally distributed species, the combination of synchronous rotation and strong atmospheric winds (such as super-rotation or day-to-night transport) produces distinct net line-of-sight velocity shifts at each terminator. This kinematic discrepancy manifests as a double-peak feature in their $K_{\rm p} - V_{\mathrm{sys}}$ maps, where each peak corresponds to the specific signal of one limb. Because the local temperature governs the ionisation state and line profiles on each side, varying the template temperature shifts our sensitivity between these two distinct thermal environments. As a result, the CCF preferentially maximises one terminator component over the other, yielding slightly different retrieved global $K_{\rm p}$ and RV values for the same element across different temperatures (we discuss these double-peaks in Sect. \ref{variation}).

\subsection{Single-line analysis}
Even after the aliasing correction, it is possible that not all of the remaining signals can be immediately interpreted as genuine atmospheric detections, as residual aliasing or spurious correlations may still affect the CCFs of some species, particularly for species whose templates contain fewer than five absorption lines, such as \BaII \, and \GaI. From the aliasing analysis, both \BaII \, and \GaI \, remained ambiguous even after correction. In particular, we found that \BaII \, correlates primarily with \FeI \,, while \GaI \, with both \MnI \, and \FeI.
Furthermore, with only two lines, respectively, in the HARPS-N wavelength range, \GaI \, is intrinsically more prone to correlate not only with other atomic species but also with noise if its absorption lines are weak, making \GaI \, detection more complicated compared to species having hundreds of absorption lines.
Hence, we complemented the CCF and aliasing analysis for \BaII \, and \GaI \, by inspecting their individual absorption lines directly in the transmission spectra. For each line of a species, we extracted the planetary transmission spectrum independently for every observed night, following the \cite{Wyttenbach_2015} prescriptions and removing the Rossiter-McLaughling \citep{Rossiter_1924,McLaughling_1924} and the centre-to-limb variation effects \citep[e.g.][]{Yan_2017,Borsa_2018} using the {\fontfamily{pcr}\selectfont StarRotator} software \citep{Hoeijmakers_2024_starrot}.
Then, we combined all nights to obtain a single transmission spectrum for the inspected line.
Once the transmission spectrum for each line was obtained, we fitted a Gaussian profile plus a continuum offset to measure the absorption depth, from which we derived the effective planetary radius $R_{\rm eff} = R_p \sqrt{1 + c/\delta}$, where $c$ is the amplitude of the detected signal obtained from the fit, while $\delta = (R_p/R_{\star})^2$ is the transit depth. Then, the $R_{\rm eff}$ value was used to repeat the analysis for each night by incorporating the updated radius into the Rossiter–McLaughlin and centre-to-limb variation correction, iterating the procedure as \cite{Fossati_Biassoni_2023}. The final transmission spectrum  was then converted from wavelength to velocity space using a common velocity grid, from –200\,\kms \, to +200\,\kms \, equally spaced.
We repeated this procedure for all the absorption lines of the inspected element and then we co-added in velocity all the transmitted spectra for all the absorption lines with a weighted mean.

\begin{figure}[t]
  \centering
    \includegraphics[width=1\linewidth]{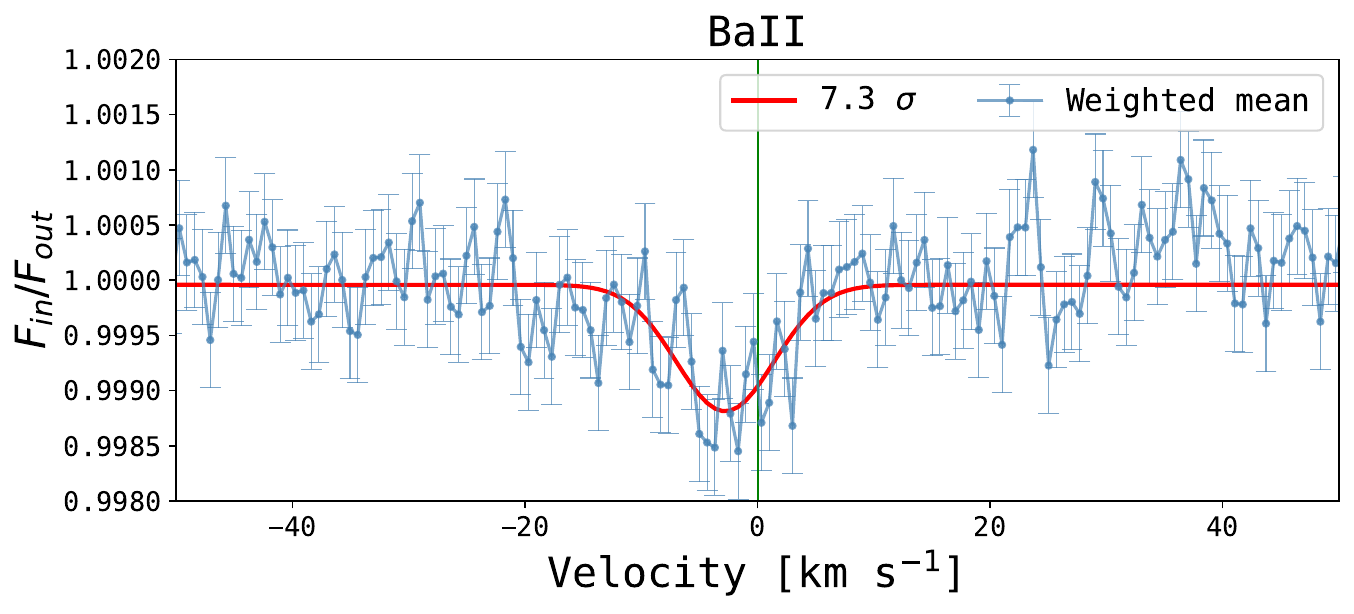}
    \caption{Transmitted spectrum of \BaII \, fitted with a Gaussian profile plus a continuum offset (solid red line) yielding a detection significance of $7.3\,\sigma$. The steel-blue points with error bars represent the combined signal from the five individual \BaII \, absorption lines, co-added across all seven nights. The vertical green line marks the expected position of the planetary signal (RV = 0\,\kms).}
    \label{fig:BaII_result}
\end{figure}

The results for \BaII \, and \GaI \, are shown in Figs. \ref{fig:BaII}, \ref{fig:GaI}, and \ref{fig:BaII_result}. We fitted a Gaussian profile plus a continuum offset (red curves) for each line and computed the significance by dividing the height of the fit by its error. For completeness, we added in the figures the template of the inspected species \BaII \, and \GaI \, (green curves) and their aliasing elements, i.e. \FeI \, (orange curves) and \MnI \, (navy curves), for each temperature from 2000\,K (light curves) to 5000\,K (dark curves) with a vertical offset.
As can be seen, the first \BaII \, absorption line (4554.033\,\AA{}) shows the strongest absorption among the five transitions and is not contaminated by \FeI \,, as illustrated in Fig. \ref{fig:BaII}. The fits to the remaining \BaII \, lines are not statistically significant, in particular for those affected by \FeI \, blending. Since only the $\sim 4554$\,\AA{} line reaches a high significance level ($\sim 5.9 \sigma$) and lies in a spectral region free of \FeI \, contamination, we conclude that the observed \BaII \, signal is genuine and not driven by \FeI \, aliasing. The weighted mean of all the \BaII \, absorption lines is depicted in the Fig. \ref{fig:BaII_result}.

\begin{figure*}[ht!]
        \centering
    \includegraphics[width=0.99\linewidth]{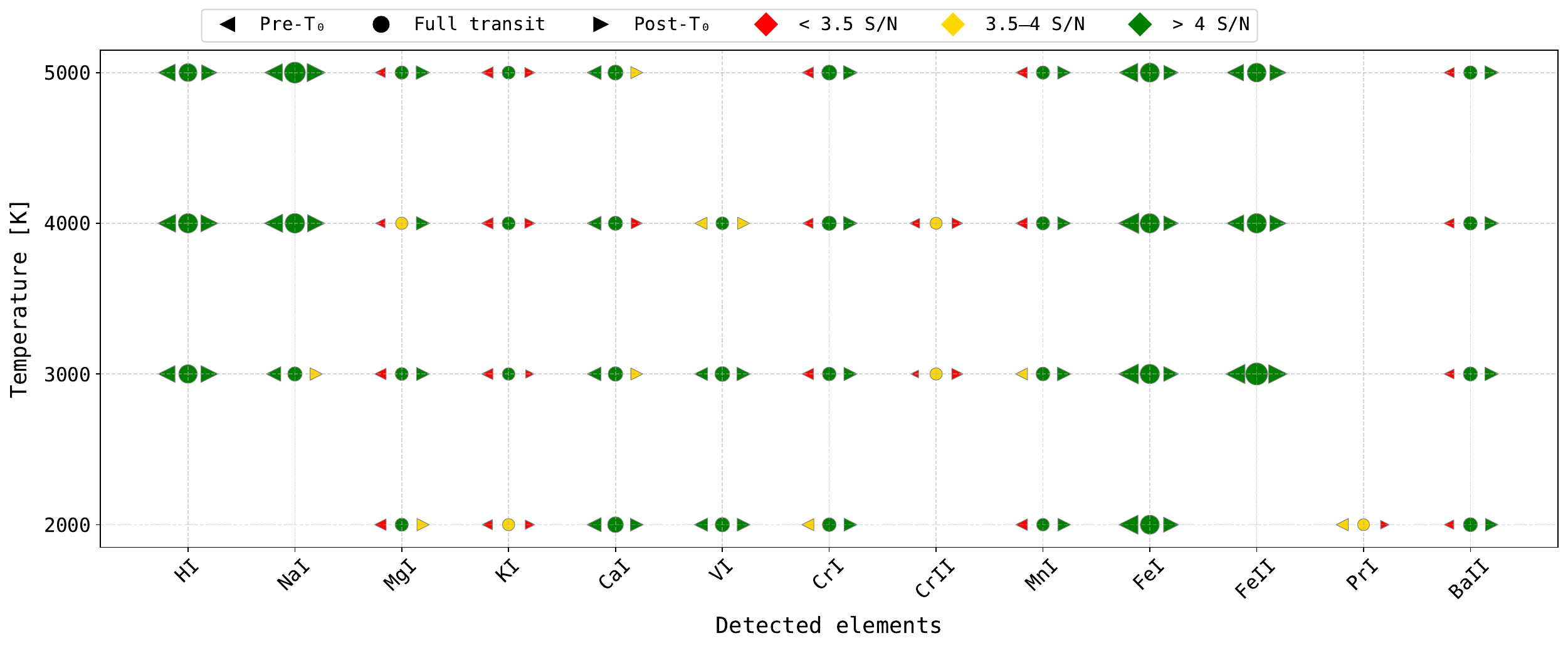}
    \caption{Results of the CCF analysis. The marker dimension corresponds to the amplitude of the Gaussian+offset fit of the $K_{\rm p}  - V_{\mathrm{sys}}$ 1D maps pre-, post-$T_0$ and full transit stacked exposure. Green markers correspond to detection with significance above the 4 S/N threshold, while yellow and red markers correspond to tentative detection (3.5 $<$ S/N $<$ 4.0) and non-detection (S/N $\leq 3.5$), respectively. The empty regions represent cases where no coherent signal was identified during the pre-selection threshold in the $K_{\rm p} - V_{\mathrm{sys}}$ maps as described in Sect. \ref{Detection criterion}}
    \label{fig:prepost}
\end{figure*}

In contrast, for \GaI \, the strongest absorption line at 4032.984\,\AA{} is significantly redshifted with respect to the expected \GaI \, position ($\sim +5$\,\kms \,, Fig. \ref{fig:GaI}). The transmission spectrum shows a significant absorption at this wavelength ($\sim 4.5 \sigma$), while the second \GaI \, line at 4172.042\,\AA{} is not statistically significant. Since the $\sim 4032$\,\AA{} feature coincides with a \MnI \, absorption line, we attribute this signal to \MnI \, rather than to \GaI. Therefore, we consider the \GaI \, detection to be spurious, due to \MnI \, contamination, and discarded the \GaI \, species from analysis.

\subsection{Looking for atmospheric longitudinal variations}\label{variation}

Given the high equilibrium temperature and the inverted pressure-temperature profile in its day-side \citep{Fossati_Biassoni_2023}, it is possible that \planet \, has a non-uniform spherical atmospheric shape but an inflated day-side compared to a smaller night-side as other ultra-hot Jupiters, for instance WASP-189b \citep[e.g.][]{Prinoth_2023}, WASP-76b \citep[e.g.][]{van_Sluijs2025}, and WASP-121b \citep[e.g.][]{Seidel_2025}.
Since ultra-hot Jupiters are thought to be tidally locked with their star, assuming this property also for \planet \, and using the parameters in Table \ref{tab:systemParameters} we obtain a changing view angle from the ingress ($T_1$ contact point) to the egress ($T_4$ contact point) of $\alpha = 2 \, \rm{arcsin} \left( \sqrt{1 - b^2} \,\, R_{\star}/a \right) \simeq 13^{\circ}$.
Hence, during the planetary transit, the stellar light is filtered through different atmospheric layers at different orbital phases. At the ingress, the stellar light is highly filtered through the leading terminator of the planet, meanwhile at the egress through the trailing terminator. Such time-dependent geometry might produce different absorption signals in our observations.

Due to the large in-transit dataset, we inspected any possible variation in time for all the detected elements similar to \cite{Lenhart_2025}. We split the merged CCFs of each species into two parts, the morning-side, i.e. from the ingress ($T_1$) to the centre of the transit $T_0$ and called it pre-$T_0$, and the evening-side, i.e. from $T_0$ to the egress ($T_4$), and called it post-$T_0$. Then, we derived their pre- and post-$T_0$ $K_{\rm p}  - V_{\mathrm{sys}}$ maps and searched for their $K_{\rm p}^{\rm max}$ values, from which we extracted the $K_{\rm p}  - V_{\mathrm{sys}}$ 1D curves.
This gave us two $K_{\rm p}  - V_{\mathrm{sys}}$ 1D curves at different $K_{\rm p}^{\rm max}$ values, one for the morning-side (pre-$T_0$) and one for the evening-side (post-$T_0$). Finally, we fitted such $K_{\rm p}  - V_{\mathrm{sys}}$ 1D curves with a Gaussian profile centred at RV $=$ 0 plus a constant offset.
Figure \ref{fig:prepost} shows all the detected and tentatively detected elements ordered by temperature and atomic number after removing the aliasing with the results of their detections of the full transit and the pre-, post-$T_0$.
The empty regions represent cases where no coherent signal was identified above the pre-selection threshold in the $K_{\rm p} - V_{\mathrm{sys}}$ maps as stated in Sect. \ref{Detection criterion}. In these instances, performing a Gaussian fit was considered both superfluous and statistically unreliable due to insufficient S/N.

Figure \ref{fig:prepost} shows that for \MgI \,, \CrI \,, \MnI \,, and \BaII \, the strongest signal belongs to the evening side compared to the morning side, where the signal is not statistically significant. The average S/N pre- and post-$T_0$ is $\sim 109 \pm 38$ and $\sim 106 \pm 42$, respectively, while the number of spectra are 116 and 113 on the morning and evening side. Hence, the highest detection observed in the post-$T_0$ phases cannot be attributed to a higher S/N or number of spectra, but is associated with a physical effect.
It is possible to explain the differences in the detected S/N of such elements considering a significantly higher atmospheric temperature in the evening region. Indeed, general circulation models of tidally locked planets predict that stellar irradiation and fast equatorial winds can transport heat eastwards and relocate the planetary hot-spot in the evening side \citep{Showman_2011}.
An elevated temperature in the evening hemisphere, as obtained from the retrieval analysis by \cite{Borsa_2022}, produces a large atmospheric scale height and increased column density at a given altitude, which in turn enhances the strength of absorption features on that side of the planet relative to the morning side. At such high temperatures, also the abundance of neutral and ion matters and might alter our detections.
The detection of \MgI \,, \CrI \,, and \MnI \, being stronger on the hotter side of the planet is somewhat surprising, as these elements ionise relatively easily, with a first ionisation energy of $\sim$\,7.65 eV, $\sim$\,6.77 eV, and $\sim$\,7.43 eV, respectively, comparable to that of \FeI \, ($\sim$\,7.90 eV). Under simple expectations, one would anticipate a reduced \MgI \,, \CrI \,, and \MnI \, abundance in hotter regions due to efficient ionisation. 
Indeed, as expected from ionisation equilibrium arguments, \FeII \, is stronger than \FeI \, in the hotter regions (Table \ref{tab:detection} and Fig. \ref{fig:prepost}). This behaviour is qualitatively consistent with temperature-driven ionisation effects.
We can attribute the lack of \MgI \,, \CrI \,, and \MnI \, in the morning hemisphere considering a cold-trap mechanism acting on the night side of the exoplanet, where these species condense and are subsequently transported by equatorial winds towards the dayside. However, by the time they reach the morning limb, they have not yet fully re-evaporated, preventing their detection in this region, while they remain observable in the evening hemisphere, which is continuously replenished by hotter gas.
To sustain this hypothesis we might look at their ionised counterparts, but within the HARPS-N wavelength range, there are too few and too weak \MgII \, and \MnII \, lines to enable a reliable detection of these ionised species. The only exception is \CrII \,, for which we obtained a tentative detection over the full transit sample. However, the S/N is not sufficient in the pre- and post-$T_0$ phases to determine whether the \CrII \, signal is preferentially associated with the morning or the evening side.

The species \MgI \,, \MnI \,, and \BaII \, also exhibit a lower $K_{\rm p}$ value compared to the theoretical one, while \CaI \, presents a higher values of significance in the morning side and $K_{\rm p}$ (Figs. \ref{fig:CCF_result} and \ref{fig:prepost}). This behaviour can be interpreted in the context of an overall asymmetric and more extended atmosphere in the evening side, subjected to stronger day-to-night side winds, where the tidally locked rotation of the planet introduces a line-of-sight velocity component that partially counteracts the orbital motion. As a result, the net projected velocity during the post-transit phases is reduced, leading to an overall decrease in the measured $K_{\rm p}$  \citep[e.g.][]{Prinoth_2022, Wardenier_2023}. On the contrary, detection obtained during the pre-$T_0$ phases could be produced by elements at lower altitudes due to a less extended atmosphere, generating a lower planetary rotation increasing the measured $K_{\rm p}$.

Analysing the first three HARPS-N nights, \cite{Nugroho_2020} found a \FeI \, double peak in their $K_{\rm p}  - V_{\mathrm{sys}}$ maps at 2000\,K, 2500\,K, and 3000\,K at different RVs and $K_{\rm p}$ values. They excluded stellar variations and different systematics as possible explanation of such behaviour. Through dedicated simulations, \citet{Nugroho_2020} showed that the \FeI \, double-peaked structure in the  $K_{\rm p}  - V_{\mathrm{sys}}$ maps arises only when two \FeI \, signals similar in $K_{\rm p}$ but shifted in RV are presented from $T_0$ to egress \citep[as shown in Fig. 16 of][]{Nugroho_2020}.
They interpreted this double-peak structure as arising from \FeI \, absorption originating from two different atmospheric limbs of the planet, which may experience different net Doppler shifts due to the combined effects of planetary rotation and atmospheric dynamics, such as equatorial jets or day–night winds. Such a double peak was also seen by \cite{Rainer_2021} using our first five HARPS-N transits.

Combining seven HARPS-N transits we used a local maxima algorithm in the $K_{\rm p}  - V_{\mathrm{sys}}$ maps to search for double peaks features. We found a \FeI \, double peak in the $K_{\rm p}  - V_{\mathrm{sys}}$ maps for all the inspected temperatures using templates different from those adopted by \cite{Nugroho_2020} and \cite{Rainer_2021}. Figure \ref{fig:FeI} shows the \FeI \, CCF results after the procedure described in the whole Sect. \ref{Data analysis} obtained with the template at 3000\,K. It is visible a double peak in the $K_{\rm p}  - V_{\mathrm{sys}}$ map at two different RVs and $K_{\rm p}$, marked by two blue crosses. Furthermore, by splitting the transit into the two temporal segments pre-$T_0$ and post-$T_0$, we were able to identify the double absorption signal in the second half of the transit, in agreement with the scenario simulated by \cite{Nugroho_2020} (Fig. \ref{fig:FeIprepost}).
Contrary to \cite{Nugroho_2020}, we obtained a similar structure also for \CrI \, (Fig. \ref{fig:CrI}). Inspecting the pre-$T_0$ and post-$T_0$ of all the detected elements, we obtained a double signal in the evening side also for \CaI \, (Fig. \ref{fig:CaIprepost}) and a double signal pre-$T_0$ for \VI \, (Fig. \ref{fig:VIprepost}). This behaviour may suggest that such species extend to different altitudes with different atmospheric dynamics that can produce Doppler-shifted signals.
To further clarify, we evaluated the velocity evolution of the \FeI \, signal for all  the inspected temperatures by tracking the position of the CCF maxima within a $\pm 10$\,\kms \, window in RV, after binning the orbital phases to enhance the S/N and preventing spurious scattered peaks.
We found a nearly symmetric central trail for all the temperatures characterised by an early redshift during ingress and a late blueshift during egress, similar to \cite{Rainer_2021}. This specific kinematic profile and double-peaks suggest that both limbs actively contribute to the absorption. Hence, the signals are dominated by planetary rotation \citep{Wardenier_2023} coupled with an eastward equatorial super-rotating jet, rather than a pure day-to-night wind as seen in WASP-76b \citep{Ehrenreich_2020}.\\

We constructed a combined template by summing the zero-normalised templates of all species taken at the temperatures that maximise the S/N. This composite template, containing all the detected and tentatively detected atmospheric species, was then used to compute the CCF over all observing nights. Before merging all the nights we corrected for the Doppler shadow each transit separately. 
We evaluated also the CCF using only the templates that exibit a double peak in the $K_{\rm p}  - V_{\mathrm{sys}}$ maps, i.e. \CaI \,, \FeI \,, \VI \,, and \CrI.

For a clearer visualisation of the phase dependence of the signal, we considered the CCFs obtained using both the templates of all species and those showing a double-peaked signature (i.e. left panels of Figs. \ref{fig:All_CCF} and \ref{fig:All_DP_CCF}). These CCFs were shifted into the planetary rest frame by adopting the theoretical orbital velocity $K_{\rm p} = 181.9$\,\kms. We then re-binned the in-transit CCFs (between the $T_1$ and $T_4$ contact points) into phase bins of 15, 18, and 20 elements, in order to reduce scatter and better highlight the phase-dependent behaviour of the signal. Since residuals of the Doppler shadow remain after rebinning, we removed them following the same procedure described in the previous steps.
For each re-binned CCF, we performed a Gaussian fit plus a constant offset (accounting for the continuum) in a velocity range centred around RV = 0 for each re-binned phase. From these fits, we derived the centre, full width at half maximum (FWHM), and amplitude of the Gaussian profiles, together with their associated uncertainties. The measured centres were then fitted using the model $v(\phi) = K\sin{(2 \pi \phi)} \,+ \, w$ allowing us to simultaneously constrain both the semi-amplitude $K$ and a potential offset $w$ in RV, where $\phi$ are the new binned phases.

We performed four separate fits: using all points between $T_1$ and $T_4$, only those between $T_2$ and $T_3$, $T_2 - T_0$, and $T_0 - T_3$.
For the CCF obtained using all templates (Fig. \ref{fig:Fit_bin_last} first row), the $K$ value derived from the $T_2 - T_3$ fit is in excellent agreement with the theoretical expectation. In contrast, the fit over the full $T_1 - T_4$ interval yields a lower $K$, likely due to the inclusion of points close to ingress and egress ($T_1 - T_2$ and $T_3 - T_4$), where the planetary signal is weaker or partially absent. Similar to \cite{Rainer_2021}, the last points show a net blue-shift, indicating an atmospheric dynamic probably caused by the strong day-to-night side wind of the evening side of the planet.
The fits restricted to the $T_2 - T_0$ and $T_0 - T_3$ consistently return slightly lower $K$ values than the theoretical one. The relatively large uncertainties in these fits are mainly due to the limited number of phase bins. We obtained these results also for the 15 and 18 bin cases.

In contrast, when considering only the species that exhibit a double-peaked signature (Fig. \ref{fig:Fit_bin_last} second row), the fitted $K$ values between $T_2$ and $T_3$ are systematically higher than the theoretical value for each bin choice, and this trend is even more pronounced for the $T_0 - T_3$ phases. The discrepancy of the $K$ values between the $T_2 - T_0$ and $T_0 - T_3$ indicates a different atmospheric geometry of the planet in these two side. This discrepancy may also be associated with differences in the temperature and pressure profiles between the two limbs, which can confine species at different atmospheric altitudes, ultimately producing a time-variable signal, as shown by the presence of a double peak structure in some species and the lowest or highest significance of our detected elements between the pre- and post-$T_0$ phases. 
In addition to the variations in $K$, the fits systematically yield a negative velocity offset, with $w \sim -2$\,\kms \, in most cases, and reaching values as low as $-3$ to $-4$\,\kms \, for the $T_2 - T_0$ phases, indicating a net day-to-night side wind.

Overall, these analyses indicates that the observed spread in $K_{\rm p}$ values is not driven by noise or template-specific systematics, but instead reflects a physical segregation among species. This suggests that different elements probe distinct regions of the planetary atmosphere, characterised by asymmetric spatial distributions and/or altitude-dependent velocity fields.

\subsection{Comparison with literature}\label{comparison}

In this work, we adopted a systemic velocity of $V_{\rm{sys}} = -24.48$\,\kms \,, calculated by \cite{Rainer_2021} based on a linear fit of the out-of-transit RVs of the first five transits reported in Table \ref{log_observation}. We note that several previous high-resolution studies of \planet \, employed slightly lower values of $V_{\rm{sys}}$ (e.g. $-21.07$\,\kms \, adopted by \citealt{hoeijmakers_2020kelt} , $-22.06$\,\kms \, by \citealt{Nugroho_2020}, and $-22.78$\,\kms \, by \citealt{Lenhart_2025}). As a consequence, the absolute RV positions of the detected planetary signals may appear offset with respect to those reported in the literature. These differences with previous works do not reflect intrinsic physical variations, such as atmospheric winds, but likely arise solely from the choice of reference frame.

Compared to previous literature results, we almost doubled the number of detections in the \planet \, atmosphere, from 7 to 13.
We compared the RV centroids of the detected species with those reported in the literature, taking into account the different values of the systemic velocity adopted in previous works.
To ensure a consistent comparison, all measurements were referred to the same absolute velocity frame by removing the corresponding $V_{\mathrm{sys}}$ used in each study, and uncertainties were propagated accordingly. The agreement was then assessed by comparing the overlap of the confidence intervals at the $3\sigma$ levels for both measurements.
Our results are consistent within $3\sigma$ with the literature detections reported by \cite{Lenhart_2025}, \cite{Nugroho_2020}, and \cite{hoeijmakers_2020kelt} for all species, except for \FeI. In this case, our measurements are inconsistent with those of \cite{Nugroho_2020} and \cite{Lenhart_2025}, but remain consistent within $3\sigma$ with \cite{hoeijmakers_2020kelt}.
Concerning the $K_{\rm p}$ values, all detected species are consistent within $3\sigma$ with the values reported by \citet{hoeijmakers_2020kelt}, \citet{Nugroho_2020}, and \citet{Lenhart_2025} but again with the only exception of \FeI \, which is consistent within 3$\sigma$ only with \cite{Lenhart_2025}.

From their injection retrieval test, \cite{Nugroho_2020} were able to recover the \VI \, signal, associating it with the possibility of thermal dissociation of vanadium oxide  (VO). However, they did not detect any V-bearing species in the HARPS-N data, attributing such non-detection to non-chemical equilibrium mechanisms, such as a vertical or night-side cold trap.
Adding four additional transits, we were able to detect the signals of \VI \, since it increased in significance, transitioning from non-detections to clear detections. Our \VI \, detection is in agreement with the injection recovery test made by \cite{Nugroho_2020}, supporting the hypothesis of VO dissociation and refuting the V cold trap. 

Adding four further nights with respect to \cite{Nugroho_2020}, the S/N should enhance the detectability of atmospheric species, similarly to what is observed for newly detected elements and \VI. In contrast, no such improvement is observed for \TiI. As also supported by \citet{Nugroho_2020}, their injection–retrieval tests predict that \TiI \, should be detectable in the data; however, no significant signal is recovered. Even after adding four more transits, \TiI \, remains undetected. This strongly suggests that titanium is depleted from the gas phase, likely due to condensation into TiO and confined by cold trap processes. Further support for this interpretation comes from our search for TiO and VO using the \cite{Kitzmann_2023} model templates, which did not yield any significant detection, indicating that TiO may be efficiently condensed, possibly on the planetary nightside, while VO may be thermally dissociated, consistent with the detection of atomic \VI.

\section{Conclusions}\label{conclusion}

For each detected species, we retained only the detection and tentative detection corresponding to the temperature that maximises the full-transit S/N. The resulting full transit S/N values, together with the corresponding pre-$T_0$ and post-$T_0$ S/Ns evaluated at the same optimal temperature, are reported in Table \ref{tab:detection}.
Table \ref{tab:detection} also summarises the kinematic properties of the detected signals, listing the centre of the Gaussian fit to the $K_{\rm p} - V_{\mathrm{sys}}$ 1D curves and the $K_{\rm p}$ mean values obtained from the bootstrap with their standard deviations.
Although the pre- and post-$T_0$ S/Ns may be higher at different temperatures, we chose to report those obtained at the temperature that maximises the full-transit signal, ensuring internal consistency for each species. Figure \ref{fig:result_all_detection} shows all the $K_{\rm p} - V_{\mathrm{sys}}$ maps computed over the full-transit duration for all detected and tentatively detected species, each extracted at the temperature that maximises the S/N.

Exploiting seven transits observed with the high-resolution HARPS-N spectrograph at the Telescopio Nazionale Galileo, we conducted a detailed atmospheric study of \planet \, using the CCF technique. By analysing multiple atomic and ionic species across a range of temperatures and correcting for the Doppler shadows and aliasing effects, we obtained robust and unambiguous detections of the following species: \HI \,, \NaI \,, \MgI \,, \KI \,, \CaI \,, \VI \,, \CrI \,, \MnI \,, \FeI \,, \FeII \,, \BaII, and tentatively detected \CrII \, and \PrI \, doubling the previously detected number of species, where \KI \,, \CaI \,, \VI \,, \MnI \,, \BaII \,, and \PrI \, are detected for the first time in the \planet \, atmosphere. Using the aliasing and single line analysis, we were able to remove spurious signals of \GaI \,, \GeI \,, \RbI \,, \DyI \,, and \TmII. In contrast with previous literature detections, thanks to a larger dataset we were able to detect \VI \, while not detecting \TiI \,, VO, or TiO. This may suggest that \TiI \, condenses into TiO on the night side of the planet, making it undetectable during transit, whereas VO may be fully dissociated, allowing for the detection of \VI.
Thanks to the large dataset, we were able to split the transit into pre- and post-mid-transit phases without significant loss of signal. We find that many species exhibit significantly stronger absorption during the post-$T_0$ phase, consistent with atmospheric asymmetries driven by higher temperatures and an inflated scale height on the evening terminator compared to the morning side. 
Furthermore, the analysis of the measured $K_{\rm p}$ values suggests that different species probe different atmospheric altitudes and experience distinct global wind patterns, leading to signals at different $K_{\rm p}$ values.
We detected a double-peaked signal in the \CaI \,, \VI \,, \CrI \,, and \FeI \, CCF and $K_{\rm p}  - V_{\mathrm{sys}}$ maps, in agreement with literature detections. The presence of double peaks indicates not only that these species probe distinct atmospheric altitudes, but also their absorption signals originate from both planetary limbs.
Overall, our results show the power of combining aliasing-aware CCF analysis with phase-resolved transmission spectroscopy to probe longitudinal variations in atmospheric composition and dynamics, paving the way for future studies capable of delivering unprecedented insights into the three-dimensional structure of exoplanet atmospheres.

\section{Data availability}
Public \planet \, reduced HARPS-N data can be found at the TNG archive \url{http://archives.ia2.inaf.it/tng/}. Private data are available from the authors upon reasonable request.

\begin{acknowledgements}
The authors acknowledge financial contribution from the INAF GO Large Grant 2023 GAPS-2, from the INAF GO Large Grant 2024 BRIDGES as well as from the European Union - Next Generation EU RRF M4C2 1.1 PRIN MUR 2022 project 2022CERJ49 (ESPLORA). FBo acknowledges support from Bando Ricerca Fondamentale INAF 2023. We acknowledge the Italian center for Astronomical Archives (IA2, \url{https://www.ia2.inaf.it}), part of the Italian National Institute for Astrophysics (INAF), for providing technical assistance, services and supporting activities of the GAPS collaboration. We acknowledge the anonymous referee for their valuable comments and suggestions.
\end{acknowledgements}

\bibliographystyle{aa}
\bibliography{bibliography.bib}

\begin{appendix}

\section{}

\begin{figure}[ht!]
    \centering
    \includegraphics[width=2\columnwidth]{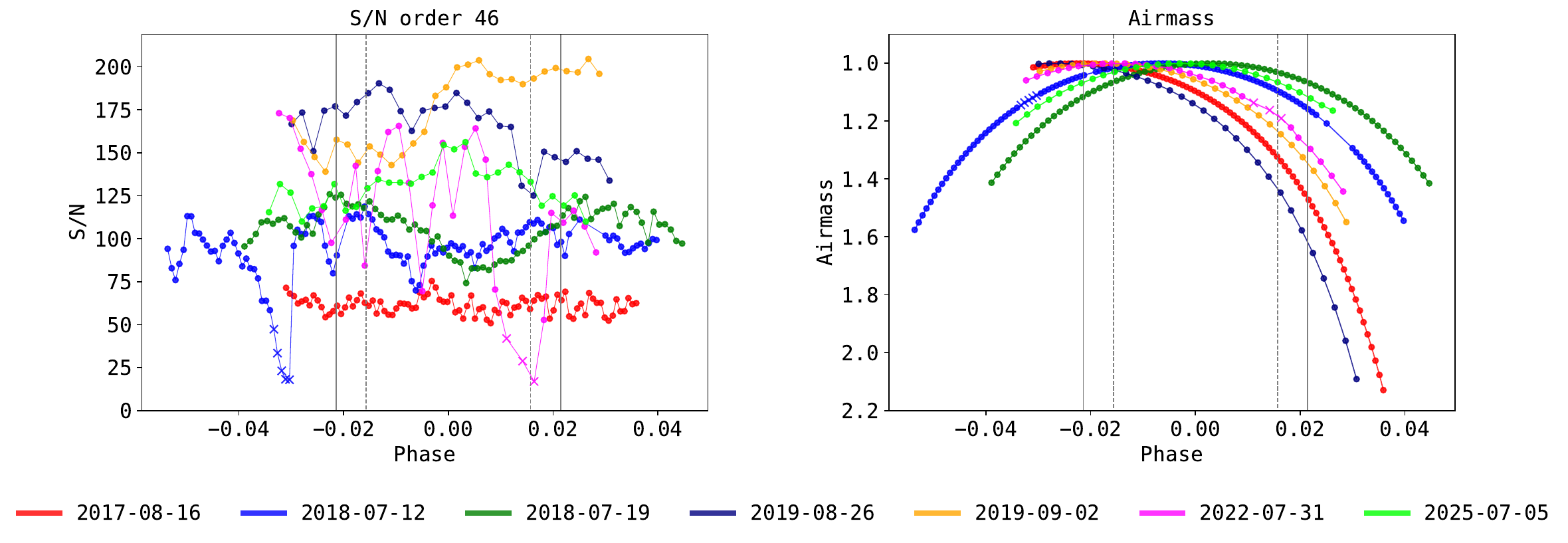}
    \begin{minipage}{\textwidth}
    \caption{S/N evaluated at order 46 (near the centre of HARPS-N band where the S/N reaches his maximum values) and airmass as a function of phase for all the nights. The S/N values below 50 are represented with cross markers and are rejected from the analysis. Vertical solid grey lines identify the beginning ($T_1$ contact point) and the end ($T_4$ contact point) of the transit, while the dashed lines mark the $T_2$ and $T_3$ contact points.}
    \label{airmass_snr}
    \end{minipage}
\end{figure}

\begin{figure}[ht!]
    \centering
    \includegraphics[width=2\columnwidth]{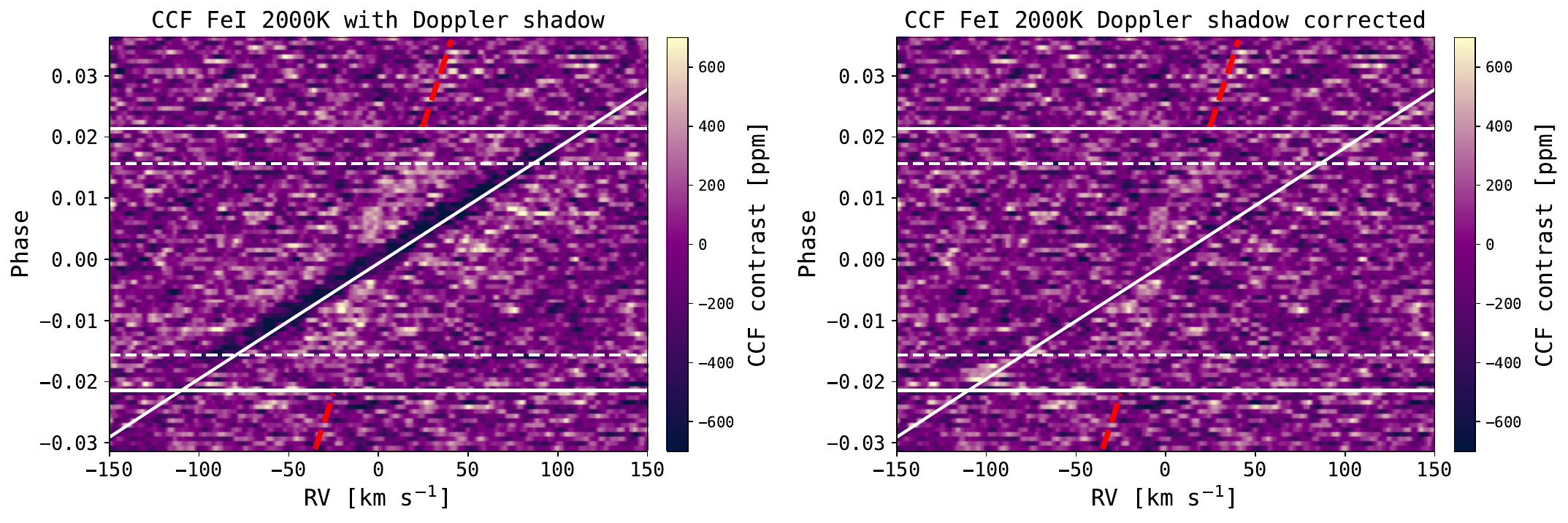}
    \begin{minipage}{\textwidth}
    \caption{Cross-correlation obtained with the \FeI \, template at 2000\,K for the first night (16 August 2017). The two horizontal solid white lines correspond to the ingress ($T_1$) and egress ($T_4$), while the dashed ones to the $T_2$ and $T_3$ contact points. The dashed red line marks the planet’s RVs. The left panel corresponds to the CCF results, where the Doppler shadow is still uncorrected. The slanted white line maps the Doppler shadow RVs according to Eq. \ref{eq:vds}. In the right CCF, the Doppler shadow is removed using a double-Gaussian fit. The colour bar range was divided by 3 to increase the contrast and to give a better visualisation of the planetary and Doppler shadow traces.}
    \label{DS}
    \end{minipage}
\end{figure}

\begin{figure*}
    \centering
    \includegraphics[width=0.79\linewidth]{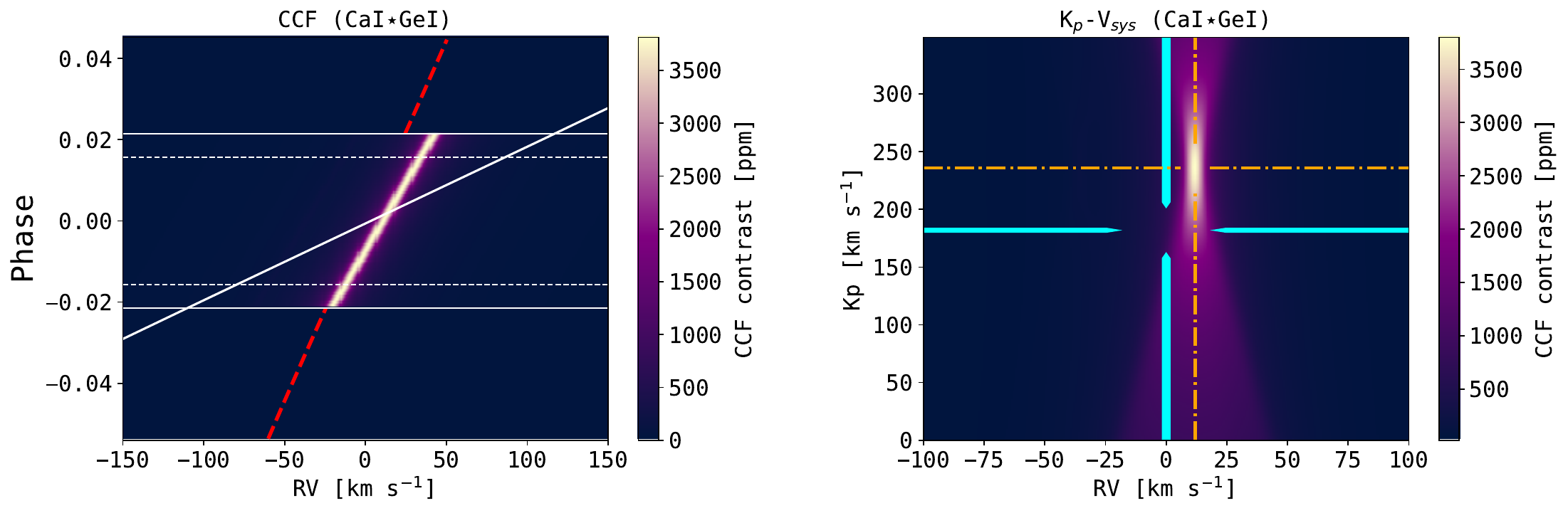}
    \caption{Example of aliasing between \GeI \, and \CaI \, at 5000\,K. (Left) CCF and (Right) $K_{\rm p}  - V_{\mathrm{sys}}$ maps of \GeI \, and \CaI \, where \GeI \, is taken as template $A$ while \CaI \, mimics the planetary signal during transit.
    The aliasing between \GeI \, and \CaI \, is visible as a white-violet signal between the ingress and egress and it is slightly more inclined than the planetary signal since the \CaI \, was Doppler shifted, assuming $K_{\rm p}$ = 236\,\kms \, which was the $K_{\rm p}^{\mathrm{max}}$ found for the \GeI \, instead of the theoretical $K_{\rm p}$ value. The cyan arrows in the $K_{\rm p}  - V_{\mathrm{sys}}$ map indicate the theoretical $K_{\rm p}$ value and the RV = 0 point. The dash-dotted gold lines aim at the maximum value of the map.}
    \label{fig:aliasing}
\end{figure*}

\newpage

\begin{figure*}[ht!]
\centering
    \includegraphics[width=0.8\linewidth]{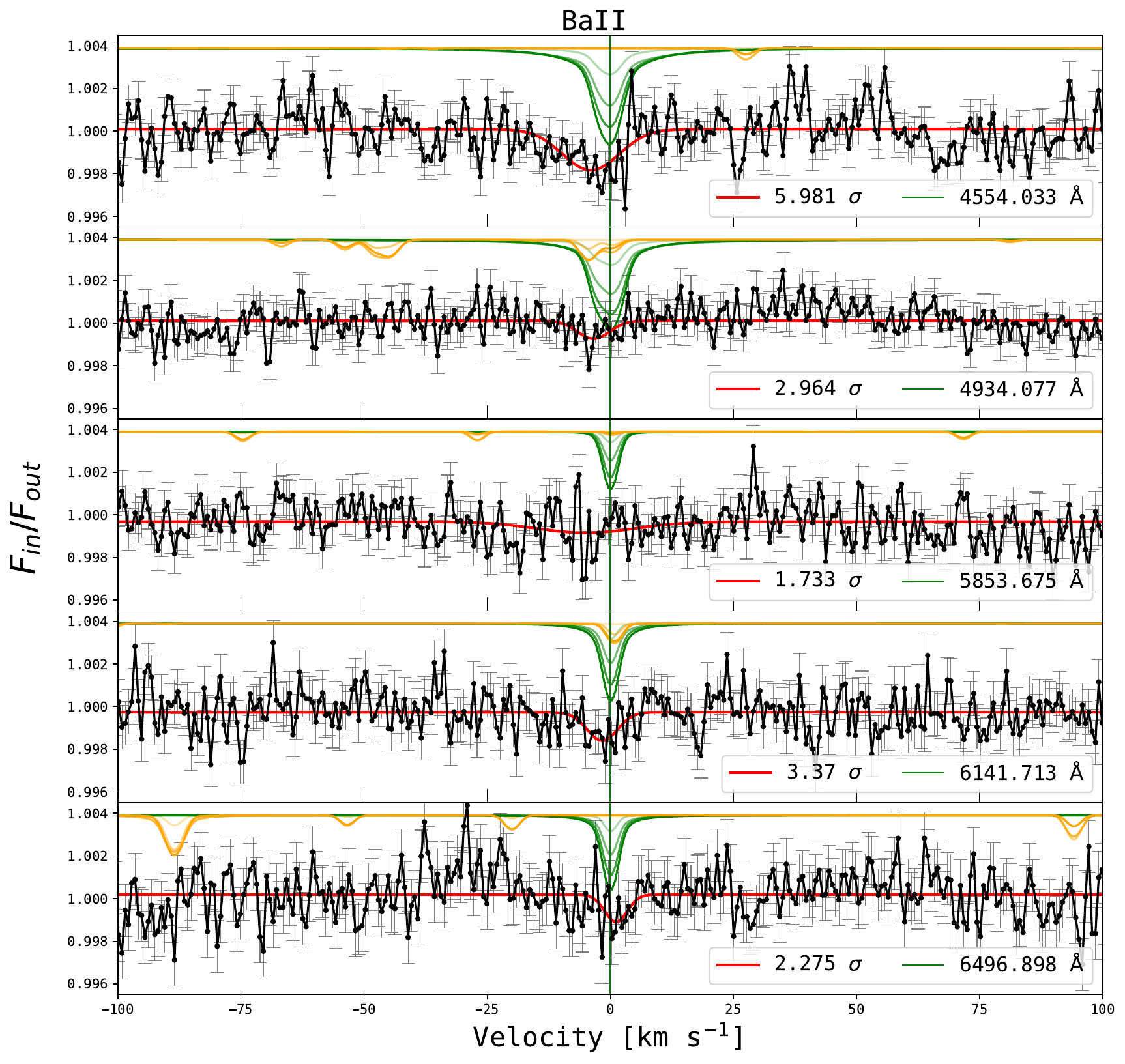}
    \caption{Analyses of \BaII \, single lines. The rows correspond to the transmitted spectra for each \BaII \, absorption line. The red curves represent the Gaussian fit of the data with their significances in the legend. Green curves represent the \BaII \, template at different temperatures, from 2000\,K (light curve) to 5000\,K (dark curve). The same holds for the \FeI \, template, represented by the orange curves. All the templates are shifted vertically by a common offset for better visualisation. The vertical green line centred at 0\,\kms \, represents the theoretical wavelength of the \BaII \, absorption lines, whose values are reported in the legend.}
    \label{fig:BaII}
\end{figure*}

\begin{figure*}[ht!]
    \centering
    \includegraphics[width=0.67\linewidth]{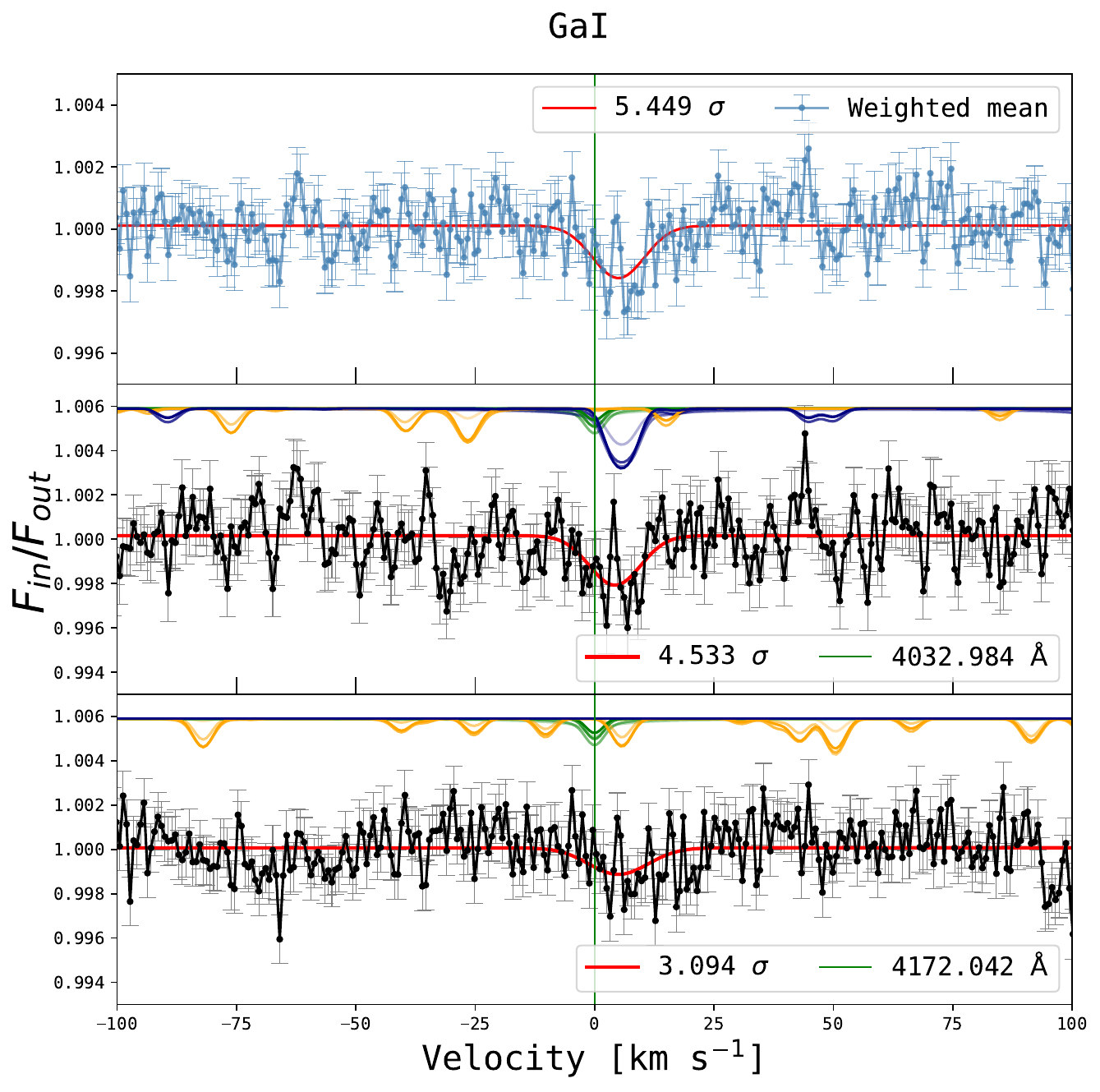}
    \caption{Same as Fig. \ref{fig:BaII} but for \GaI. The first row corresponds to the weighted mean of the absorption spectra. The \GaI \,, \MnI \,, and \FeI \, templates are represented in green, navy, and orange, respectively. It is visible how the strongest absorption at 4032.984\,\AA{} is due to \MnI \, rather than \GaI. Hence, we considered its detection to be spurious due to \MnI \, contamination, and discarded \GaI \, from our analysis.}
    \label{fig:GaI}
\end{figure*}

\begin{figure*}[ht!]
    \centering
    \includegraphics[width=0.9\linewidth]{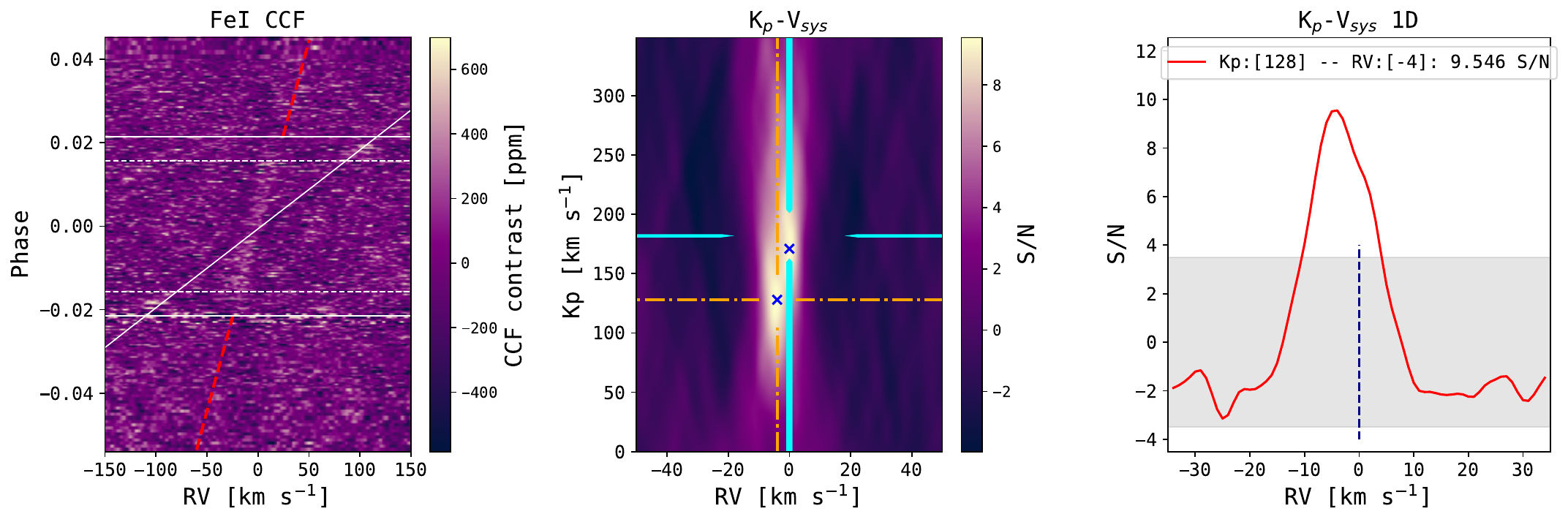}
    \caption{CCF and $K_{\rm p}  - V_{\mathrm{sys}}$ maps of \FeI \, obtained with the template at 3000\,K. The first figure is the CCF map. The slanted dashed red line corresponds to the planetary RVs, while the slanted white line follows the Doppler shadow. The two solid horizontal white lines correspond to the $T_1$ and $T_4$ contact points, while the dashed ones to the $T_2$ and $T_3$ contact points. The colour bar range was divided by 3 to increase the contrast of the planetary trace in the CCF map. The second panel is the $K_{\rm p}  - V_{\mathrm{sys}}$ map. The cyan arrows point at the expected planetary signal centred in RV $=$ 0\,\kms \, and the theoretical $K_{\rm p} \simeq$ 181.9\,\kms. The dash-dotted orange lines aim at the minimum value of the map. The last figure is the $K_{\rm p}  - V_{\mathrm{sys}}$ 1D curve. It was evaluated at the minimum $K_{\rm p}^{\rm min}$ value found in the $K_{\rm p}  - V_{\mathrm{sys}}$ map. The navy vertical line is centred at RV $=$ 0\,\kms \,, while the grey region is the 3.5 S/N level. In legend are shown the $K_{\rm p}^{\rm min}$ and RV values of the minimum pixel of the $K_{\rm p}  - V_{\mathrm{sys}}$ map with its significance. The two blue crosses mark the two peaks with S/N $> 8.5$.}
    \label{fig:FeI}
\end{figure*}

\begin{figure*}[ht!]
    \centering
    \includegraphics[width=1\linewidth]{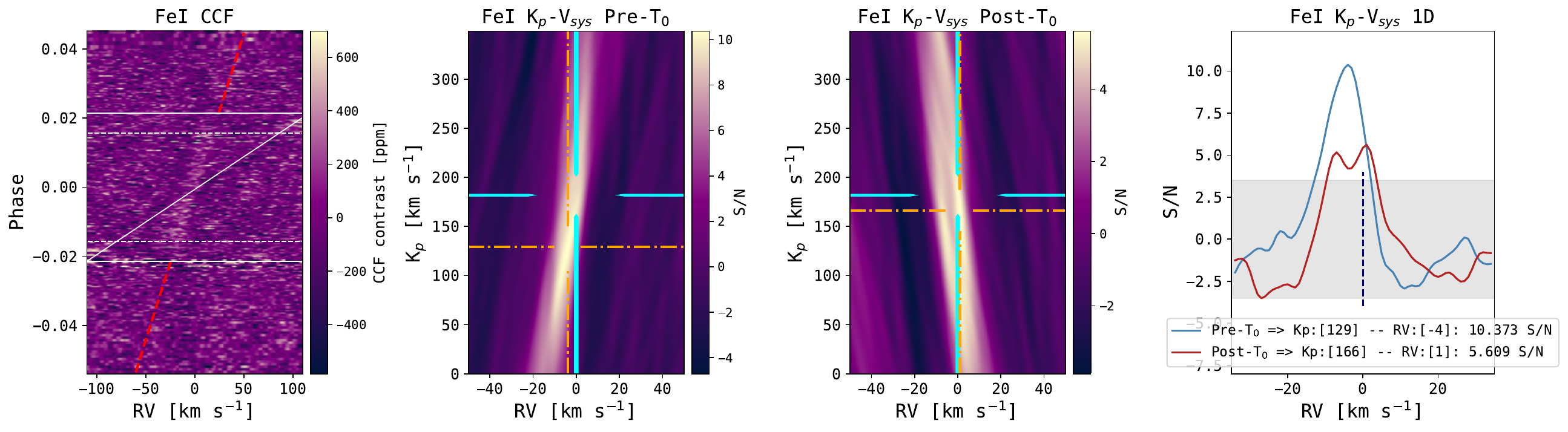}
    \caption{CCF and $K_{\rm p}  - V_{\mathrm{sys}}$ maps of \FeI \, obtained with the template at 3000\,K as in Fig. \ref{fig:FeI}. The pre-$T_0$ $K_{\rm p}  - V_{\mathrm{sys}}$ map is derived from the CCF between $T_1$ and $T_0$, while the post-$T_0$ from $T_0$ to $T_4$. The last figure shows the $K_{\rm p}  - V_{\mathrm{sys}}$ 1D curves evaluated at the minimum $K_{\rm p}^{\rm min}$ values of the two $K_{\rm p}  - V_{\mathrm{sys}}$ maps, where in its legend are shown the $K_{\rm p}^{\rm min}$ and the corresponding RV values of the minimum value with their significance. It is visible that in the evening side, the \FeI \, is composed of a double signal with similar $K_{\rm p}$ but different RV.}
    \label{fig:FeIprepost}
\end{figure*}

\begin{figure*}[ht!]
    \centering
    \includegraphics[width=0.9\linewidth]{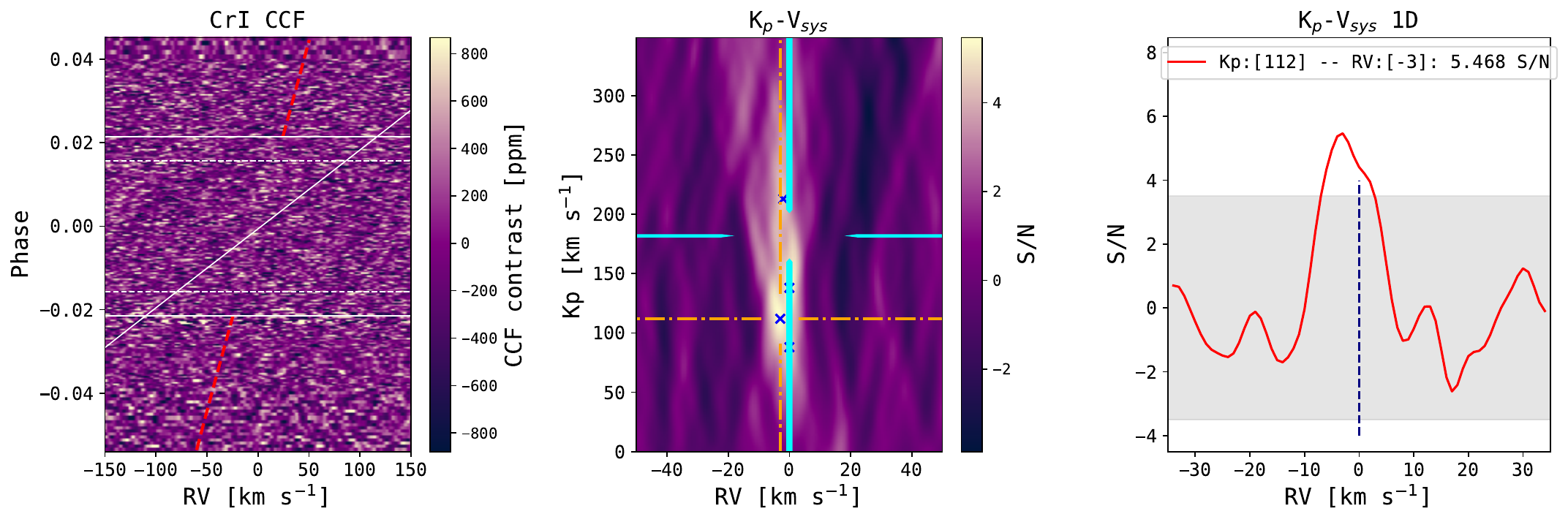}
    \caption{Similar to Fig. \ref{fig:FeI} but for \CrI \, at the inspected temperature of 5000\,K.}
    \label{fig:CrI}
\end{figure*}

\begin{figure*}[ht!]
    \centering
    \includegraphics[width=1\linewidth]{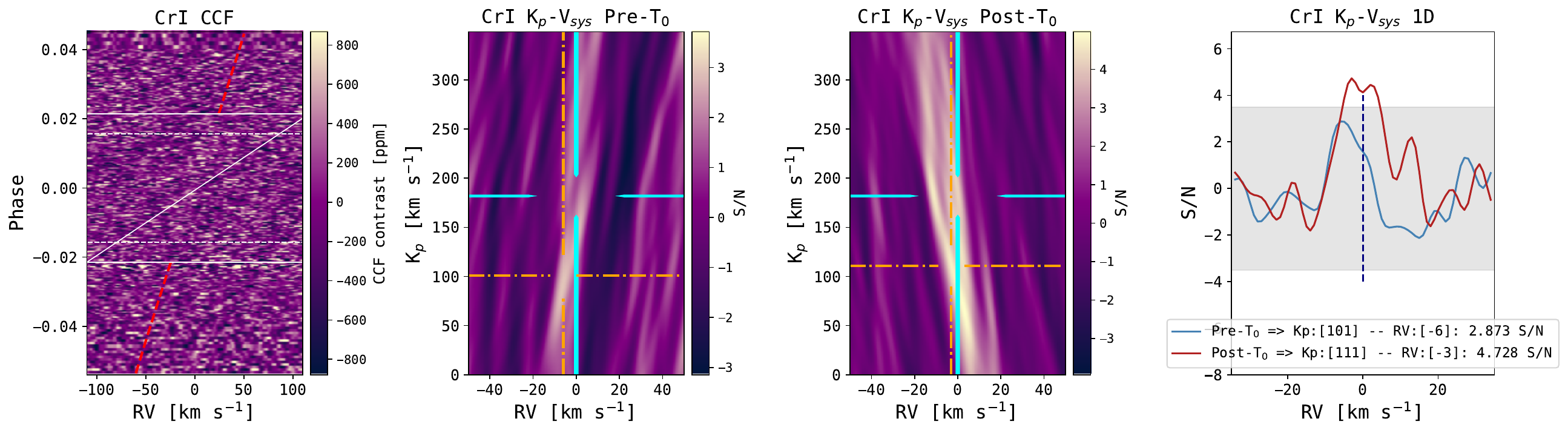}
    \caption{Similar to Fig. \ref{fig:FeIprepost} but for \CrI \, at the inspected temperature of 5000\,K.}
    \label{fig:CrIprepost}
\end{figure*}

\begin{figure*}[ht!]
    \centering
    \includegraphics[width=1\linewidth]{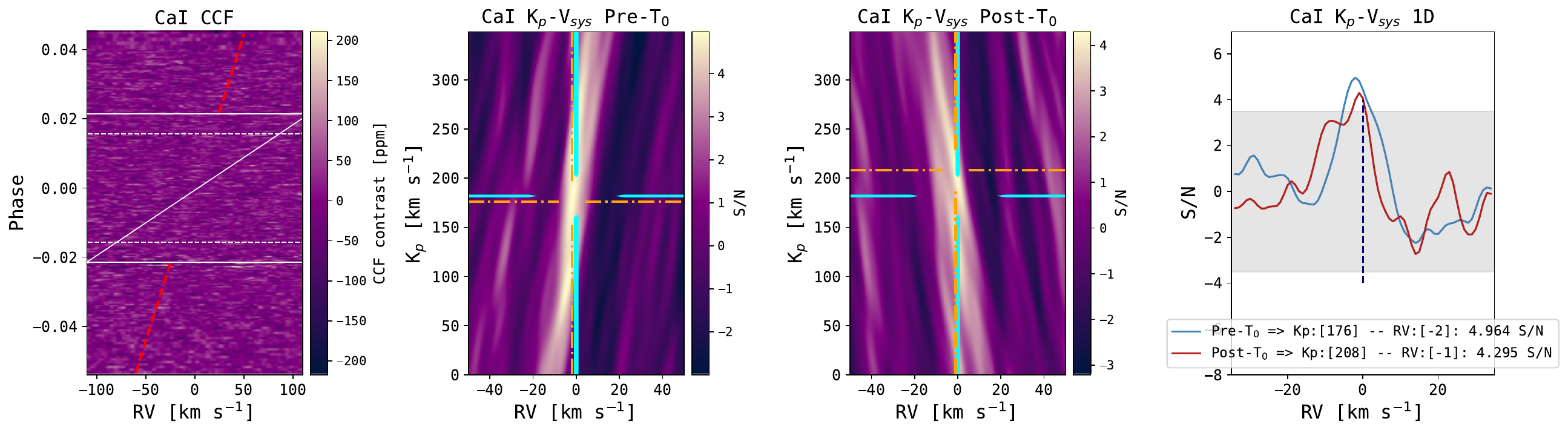}
    \caption{Similar to Fig. \ref{fig:FeIprepost} but for \CaI \, with the template at 2000\,K.}
    \label{fig:CaIprepost}
\end{figure*}

\begin{figure*}[ht!]
    \centering
    \includegraphics[width=1\linewidth]{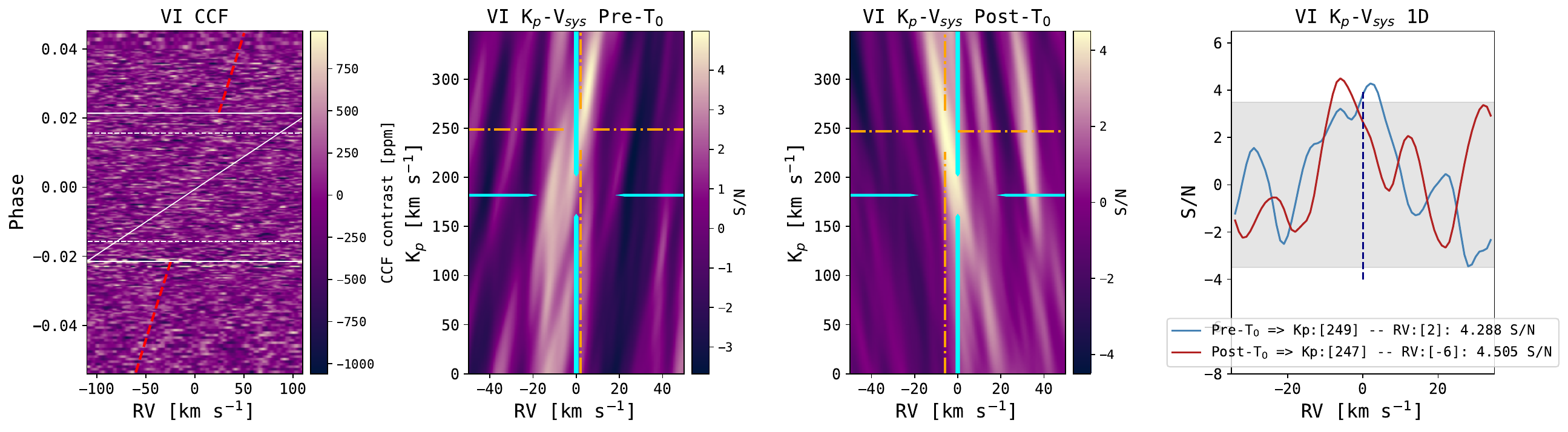}
    \caption{Similar to Fig. \ref{fig:FeIprepost} but for \VI \, with the template at 3000\,K.}
    \label{fig:VIprepost}
\end{figure*}

\begin{figure*}[ht!]
    \centering \includegraphics[width=1\linewidth]{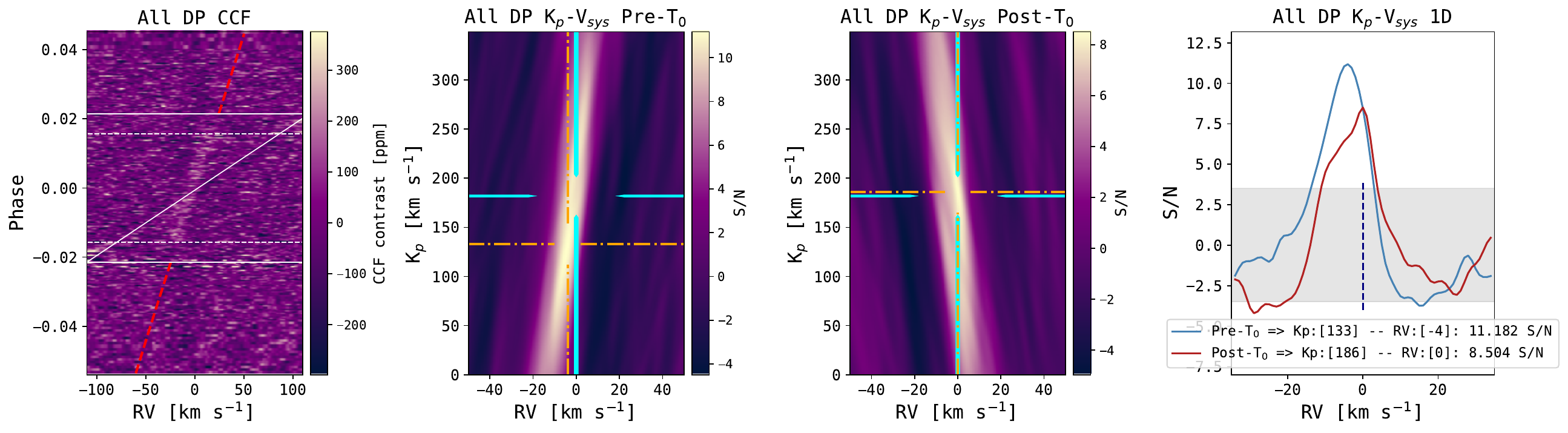}
    \caption{Similar to Fig. \ref{fig:FeI} but for all detected species having a double peak profile. We used only the templates that maximise the S/N values among different temperatures.}
\label{fig:All_DP_CCF}
\end{figure*}

\begin{figure*}[ht!]
    \centering \includegraphics[width=1\linewidth]{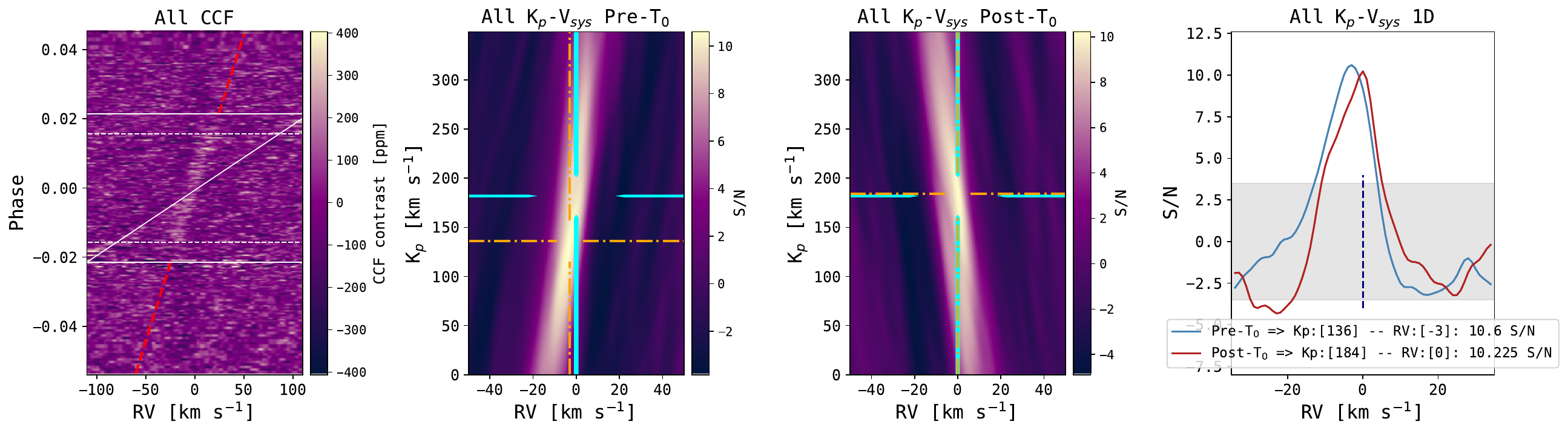}
    \caption{Similar to Fig. \ref{fig:All_DP_CCF} but for all detected species.}
\label{fig:All_CCF}
\end{figure*}

\begin{figure*}[ht!]
    \centering
    \includegraphics[width=1\linewidth]{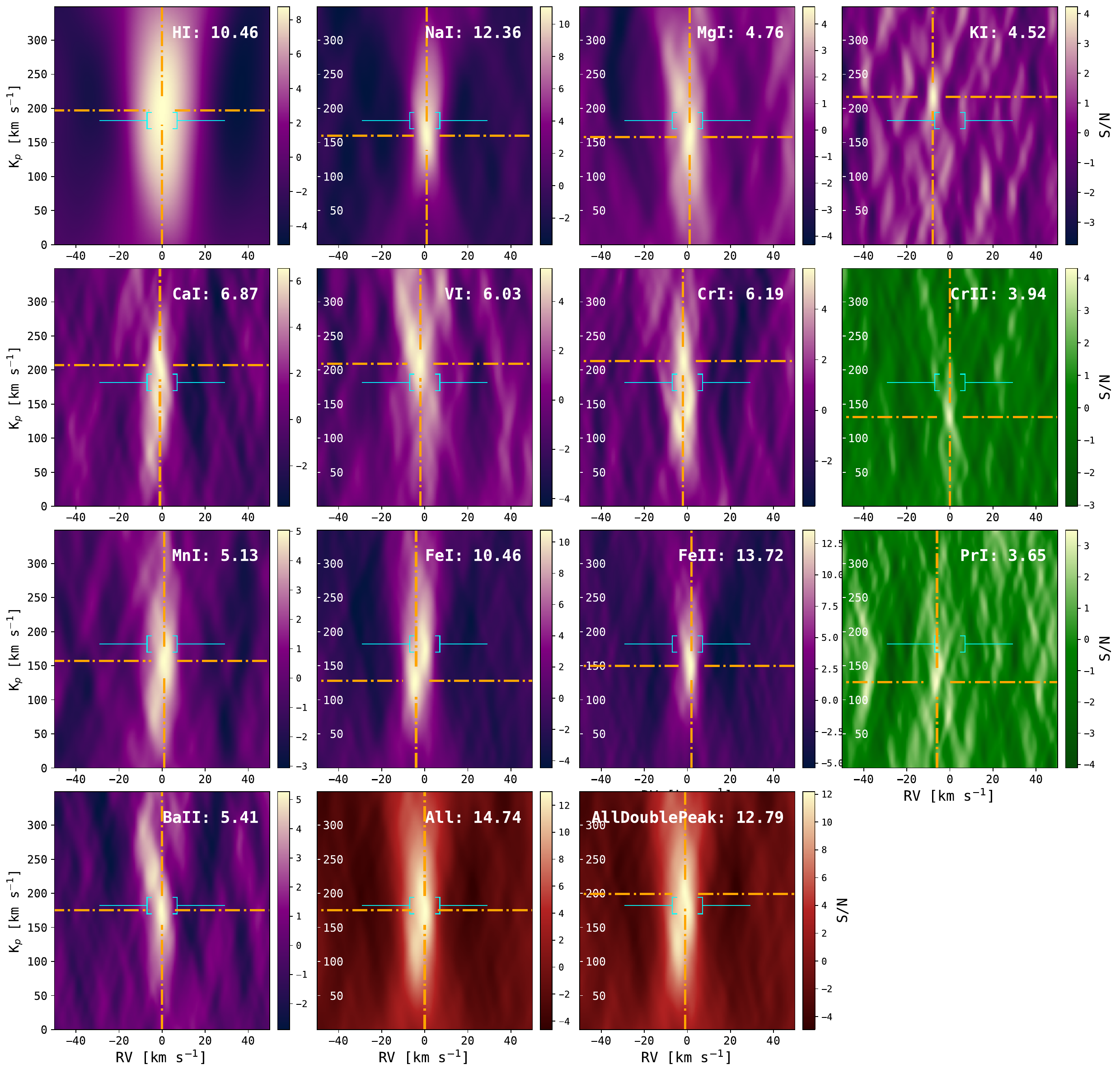}
    \caption{$K_{\rm p}  - V_{\mathrm{sys}}$ maps of all detected and tentatively detected species at the temperature that maximises the S/N. Green figures refer to our tentative detections. The cyan sight aims at the expected signal i.e. RV = 0\,\kms \, and $K_{\rm p}$ = 181.9\,\kms. Orange lines point to the maximum value of the maps, while in the text we report the S/N values obtained from the Gaussian fit of the corresponding 1D $K_{\rm p}  - V_{\mathrm{sys}}$ maps evaluated at $K_{\rm p}^{\rm max}$.}
\label{fig:result_all_detection}
\end{figure*}

\begin{figure*}[ht!]
    \centering

    \begin{subfigure}[b]{\linewidth}
    \centering
\includegraphics[width=0.9\linewidth]{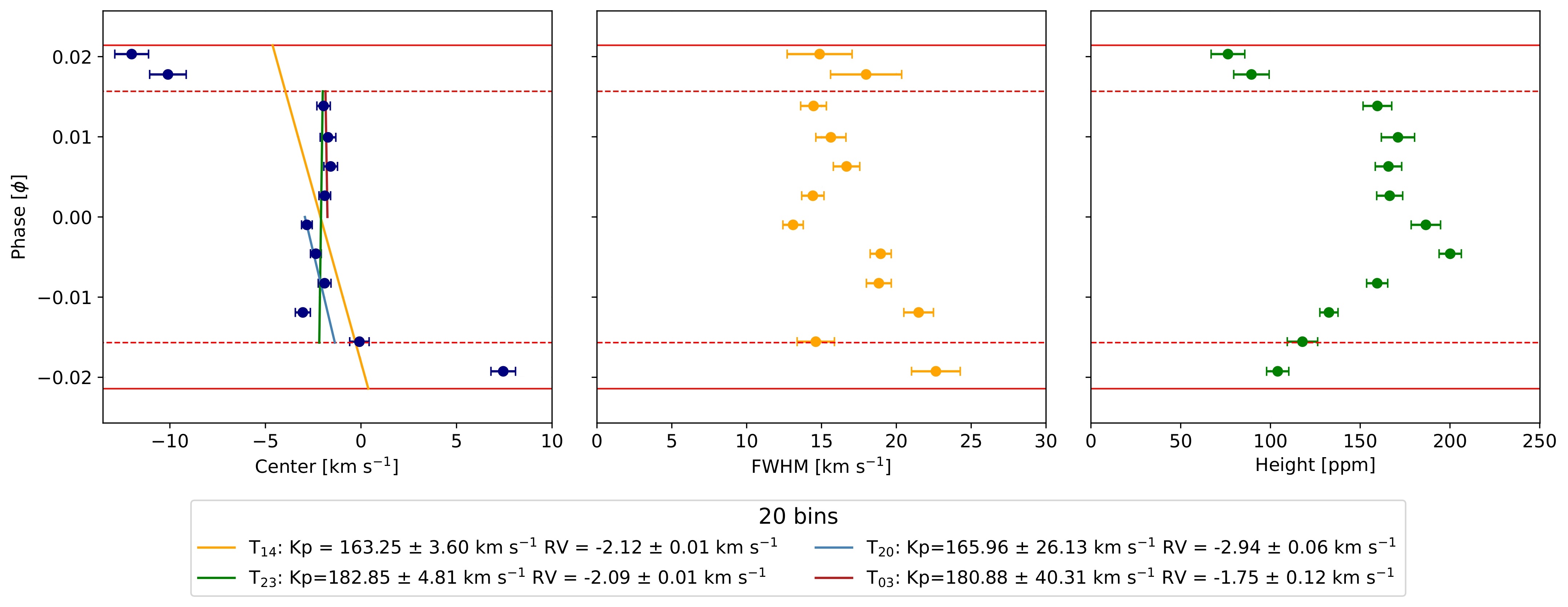}

    \end{subfigure}

    \par\vspace{0.5cm}

    \begin{subfigure}[b]{\linewidth}
        \centering
        \includegraphics[width=0.9\linewidth]{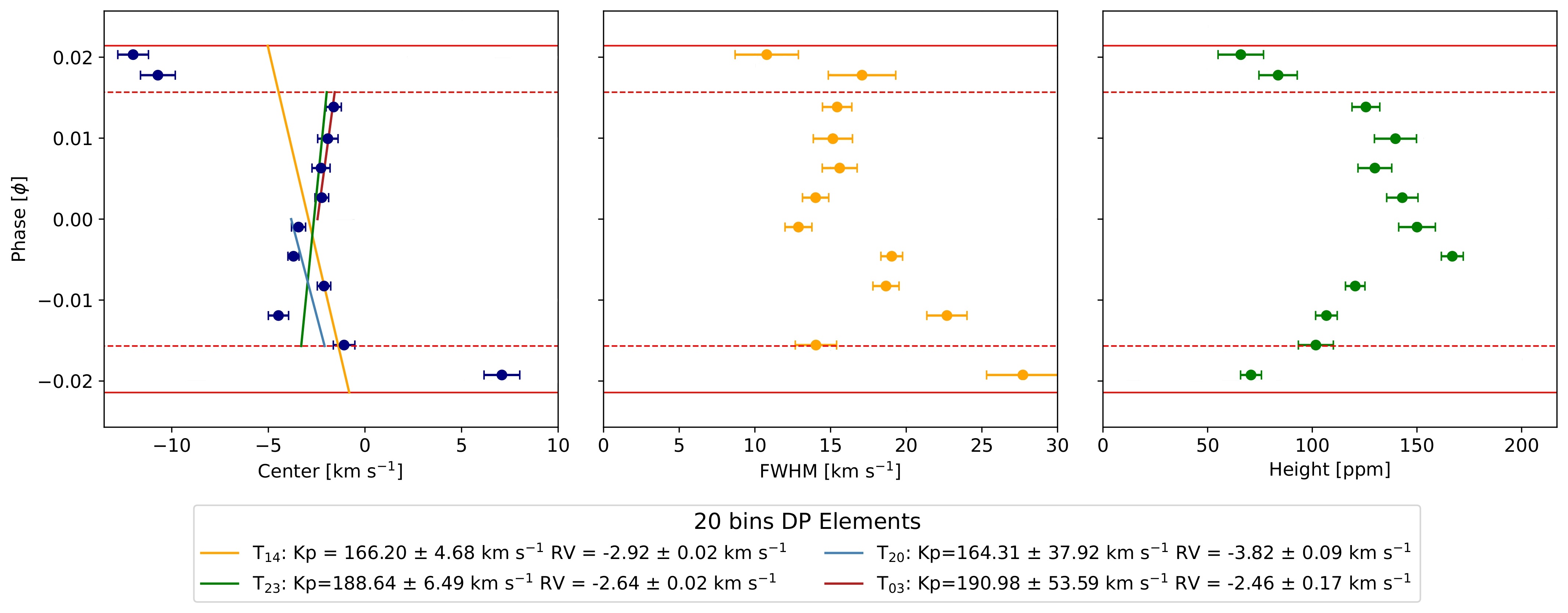}
    \end{subfigure}
    \caption{Gaussian fits to the re-binned CCFs and 20 bins. First row corresponds to the CCF obtained with all species. The first panel shows the Gaussian centre derived by the Gaussian $+$ offset fit with the results of the planetary model curve; the orange, green, blue, and red lines correspond to the model between $T_1 - T_4$, $T_2 - T_3$, pre-$T_0$, and post-$T_0$, respectively. The middle and right panels show the FWHM and the height of the Gaussian fit, respectively. The dotted and solid horizontal red lines corresponds to the four contact points.
    Second row: Same as first row, but for the CCF evaluated with the species presenting a double peak profile.}
    \label{fig:Fit_bin_last}
\end{figure*}

\begin{table*}
    \centering
    \renewcommand{\arraystretch}{1.1}
    \fontsize{10}{13.5}\selectfont
    \caption{Gaussian fit parameters and $K_{\rm p}$ values for the detected species.}
    \begin{tabular}{lcccccc}
    \hline
    \hline
     Element & Full transit S/N & RV [\kms] & $K_{\rm p}$ [\kms] & Pre-$T_0$ S/N & Post-$T_0$ S/N & New detection \\
    \hline
    \HI & 10.46 $\pm$ 0.43 & $0.45 \pm 0.47 $ &  $196 \pm 34$ & 8.91 $\pm$ 0.37 & 8.16 $\pm$ 0.29 & X\\  
    \NaI & 12.36 $\pm$ 0.49 & $0.93 \pm 0.22$  & $161 \pm 11$ & 8.68 $\pm$ 0.88 & 9.36 $\pm$ 0.54 & X\\
    \MgI & 4.76 $\pm$ 0.46 & $0.34 \pm 0.58$ & $146 \pm 24$ & 2.46 $\pm$ 0.34 & 5.02 $\pm$ 0.60 & X\\  
    \KI & 4.52 $\pm$ 0.63 & $-7.60 \pm 0.31$ & $215 \pm 14$ & 3.33 $\pm$ 0.78 & 2.71 $\pm$ 0.82 & \checkmark\\   
    \CaI & 6.87 $\pm$ 0.46 & $-1.80 \pm 0.36$ & $194 \pm 31$ & 5.39 $\pm$ 0.45 & 4.32 $\pm$ 0.43 & \checkmark\\
    \VI & 6.03 $\pm$ 0.41 & $-2.48 \pm 0.44$ & $208 \pm 15$ & 4.53 $\pm$ 0.38 & 4.90 $\pm$ 0.54 & \checkmark\\    
    \CrI & 6.19 $\pm$ 0.43 & $-2.15 \pm 0.33$ & $142 \pm 30$ & 3.14 $\pm$ 0.51 & 5.37 $\pm$ 0.34 & X\\
    \CrII & 3.94 $\pm$ 0.67 & $-0.30 \pm 0.33$ & $133 \pm 14$ & 1.29 $\pm$ 0.47 & 3.29 $\pm$ 0.75 & X \\   
    \MnI & 5.13 $\pm$ 0.35 & $0.28 \pm 0.46$ & $155 \pm 24$ & 3.61 $\pm$ 0.40 & 4.94 $\pm$ 0.30 & \checkmark\\    
    \FeI & 10.46 $\pm$ 0.47 & $-3.43 \pm 0.26$ & $128 \pm 13$ & 11.39 $\pm$ 0.42 & 6.02 $\pm$ 0.54 & X\\   
    \FeII & 13.72 $\pm$ 0.81 & $0.93 \pm 0.22$  & $172 \pm 13$ & 10.41 $\pm$ 0.82 & 9.72 $\pm$ 0.59 & X \\    
    \PrI & 3.65 $\pm$ 0.84  & $-6.47 \pm 0.44$ & $137 \pm 19$ &  3.87 $\pm$ 0.73 & 1.70 $\pm$ 0.98 & \checkmark \\
    \BaII & 5.41 $\pm$ 0.33 & $-1.19 \pm 0.32$ & $172 \pm 15$ & 2.32 $\pm$ 0.34 & 4.84 $\pm$ 0.47 & \checkmark\\
    All & 14.74 $\pm$ 0.49 & $-1.87 \pm 0.22$ & $176 \pm 18$ & 11.74 $\pm$ 0.44 & 10.69 $\pm$ 0.57 & $-$\\
    \hline
    \end{tabular}
    \tablefoot{Amplitude and centre of the Gaussian fit obtained for the full-transit pre- and post-$T_0$ for each detected element, together with the $K_{\rm p}$ value obtained from the bootstrap.}
    \label{tab:detection}
\end{table*}

\end{appendix}

\end{document}